\documentclass[11pt]{article}
\usepackage{amssymb}
\usepackage{amsmath}
\usepackage{amsmath}
\usepackage{amsthm}
\usepackage{amsfonts}
\usepackage{graphicx}
\usepackage{enumerate}
\usepackage{bbm} 
\usepackage{dsfont}
\usepackage[authoryear,longnamesfirst]{natbib}
\usepackage{multirow}
\usepackage{multicol}
\usepackage{tikz}
\numberwithin{equation}{section}
\newtheorem{theorem}{Theorem}
\newtheorem{algorithm}{Algorithm}

\newtheorem{corollary}{Corollary}

\newtheorem{lemma}{Lemma}
\theoremstyle{definition}
\newtheorem{assumption}{Assumption}
\newtheorem{example}{Example}
\newtheorem{remark}{Remark}

\DeclareMathOperator{\E}{\text{E}}

\newcommand{\supp}{\rm{supp}}
\renewcommand{\hat}{\widehat}
\renewcommand{\tilde}{\widetilde}

\newcommand{\sumi}{\sum_{i=1}^N }
\newcommand{\sumj}{\sum_{j=1}^M }
\newcommand{\Ep}{{\rm E}_{P}}
\newcommand{\Epn}{{\rm E}_{P_n}}
\newcommand{\sumij}{\sum_{i=1}^N \sum_{j=1}^M}
\newcommand{\sumkl}{\sum_{(k,\ell)\in [K]^2}}
\newcommand{\Real}{\mathbbm R}
\newcommand{\C}{\underline C}
\newcommand{\GC}{\mathbb{G}_{n}}
\newcommand{\Gnkl}{\mathbbm{G}_{n,k\ell}}
\newcommand{\G}{\mathcal{G}}

\newcommand{\T}{\mathcal T}

\newcommand{\Op}{O_P}
\newcommand{\Opn}{O_{P_n}}
\newcommand{\op}{o_P}
\newcommand{\opn}{o_{P_n}}
\renewcommand{\E}{\mathcal E}
\newcommand{\Enkl}{\mathbbm E_{n,k\ell}}
\newcommand{\sumIJ}{\sum_{(i,j)\in I_k\times J_\ell}}
\newcommand{\kk}{\emph{\textbf{k}}}
\newcommand{\sumjj}{\sum_{j_1=1}^{C_1}...\sum_{j_\ell=1}^{C_\ell}}
\newcommand{\sumk}{\sum_{\emph{\textbf{k}}\in[K]^\ell}}
\newcommand{\Enk}{\mathbbm E_{n,\emph{\textbf{k}}}}
\newcommand{\prodI}{|I_\kk|^2}
\newcommand{\prodll}{|I_\kk|}
\newcommand{\Gnkk}{\mathbbm{G}_{n,\kk}}
\newcommand{\sumN}{\sum\limits_{\imath\in [N_\emph{\textbf{j}}]}}
\newcommand{\sumNN}{\sum\limits_{\imath\in [N_\textbf{0}]}}
\newcommand{\Enkk}{\mathbbm E_{n,\emph{\textbf{k}}}}
\newcommand{\sumjjj}{\sum\limits_{\imath'\in [N_{\emph{\textbf{j}}^{\:\prime}}]}}
\begin{document}

\title{Multiway Cluster Robust Double/Debiased Machine Learning\thanks{First arXiv date: September 8, 2019\smallskip}}
\author{
Harold D. Chiang\thanks{Harold D. Chiang: harold.d.chiang@vanderbilt.edu. Department of Economics, Vanderbilt University, VU Station B \#351819, 2301 Vanderbilt Place, Nashville, TN 37235-1819, USA\smallskip} 
\qquad Kengo Kato\thanks{Kengo Kato: kk976@cornell.edu. Department of Statistics and Data Science, Cornell University, 1194 Comstock Hall, Ithaca, NY 14853, USA\smallskip}
\qquad Yukun Ma\thanks{Yukun Ma: yukun.ma@vanderbilt.edu. Department of Economics, Vanderbilt University, VU Station B \#351819, 2301 Vanderbilt Place, Nashville, TN 37235-1819, USA\smallskip} \qquad Yuya Sasaki\thanks{Yuya Sasaki: yuya.sasaki@vanderbilt.edu. Department of Economics, Vanderbilt University, VU Station B \#351819, 2301 Vanderbilt Place, Nashville, TN 37235-1819, USA\smallskip} \thanks{We benefited from useful comments by seminar participants at Southern Methodist University, Stony Brook University, University of Bristol, and University of Colorado - Boulder, and participants at CeMMAP UCL/Vanderbilt Joint Conference on Advances in Econometrics and CeMMAP Workshop on Causal Learning with Interactions. All remaining errors are ours.\bigskip}
}
\date{}

\maketitle

\begin{abstract}
This paper investigates double/debiased machine learning (DML) under multiway clustered sampling environments. 
We propose a novel multiway cross fitting algorithm and a multiway DML estimator based on this algorithm. 
We also develop a multiway cluster robust standard error formula.
Simulations indicate that the proposed procedure has favorable finite sample performance. 
Applying the proposed method to market share data for demand analysis, we obtain larger two-way cluster robust standard errors for the price coefficient than non-robust ones in the demand model.
\bigskip\\${}$\bigskip\\${}$
\end{abstract}

\section{Introduction}
We propose a novel multiway cross fitting algorithm and a double/debiased machine learning (DML) estimator based on the proposed algorithm.
This objective is motivated by recently growing interest in use of dependent cross sectional data and recently increasing demand for DML methods in empirical research.
On one hand, researchers frequently use multiway cluster sampled data in empirical studies, such as network data, matched employer-employee data, matched student-teacher data, scanner data where observations are double-indexed by stores and products, and market share data where observations are double-indexed by market and products.
On the other hand, we have witnessed rapidly increasing popularity of machine learning methods in empirical studies, such as random forests, lasso, post-lasso, elastic nets, ridge, deep neural networks, and boosted trees among others.
To date, available DML methods focus on i.i.d. sampled data.
In light of the aforementioned research environments today, a new method of DML that is applicable to multiway cluster sampled data may well be of interest by empirical researchers.

The DML was proposed by the recent influential paper by \citet[CCDDHNR,][]{CCDDHNR18}.
They provide a general DML toolbox for estimation and inference for structural parameters with high-dimensional and/or infinite-dimensional nuisance parameters.
In that paper, the estimation method and properties of the estimator are presented under the typical microeconometric assumption of i.i.d. sampling.
We advance this frontier literature of DML by proposing a modified DML estimation procedure with multiway cross fitting, which accommodates multiway cluster sampled data.
Even for multiway cluster sampled data, we show that the proposed DML procedure works under nearly identical set of assumptions to that of CCDDHNR (\citeyear{CCDDHNR18}).
To our best knowledge, the present paper is the first to consider generic DML methods under multiway cluster sampling.

Another branch of the literature following the seminal work by \cite{CGM11} proposes multiway cluster robust inference methods.
\cite{Menzel17} conducts formal analyses of bootstrap validity under multiway cluster sampling robustly accounting for non-degenerate and degenerate cases.
\cite{DDG18} develop empirical process theory under multiway cluster sampling which applies to a large class of models.
We advance this practically important literature by developing a multiway cluster robust inference method based on DML. 
In deriving theoretical properties of the proposed estimator, we take advantage of the Aldous-Hoover representation employed by the preceding papers.
To our knowledge, the present paper is the first in this literature on multiway clustering to develop generic DML methods.
 
\subsection{Relations to the Literature}
The past few years have seen a fast growing  literature in machine learning based econometric methods. For general overviews of the field, see, e.g., \cite{AtheyImbens19} or \cite{MullainathanSpiess17}. For a review of estimation and inference methods for high-dimensional data, see \cite{BCH14review}. For an overview of data sketching methods tackling computationally impractically large number of observations, see \cite{LeeNg19}. 
The DML of CCDDHNR (\citeyear{CCDDHNR18}) is built upon \cite{BCK15}, which proposes to use Neyman orthogonal moments for a general class of Z-estimation statistical problems in the presence of high-dimensional nuisance parameters. 
This framework is further generalized in different directions by \cite{BCFH17} and \cite{BCCW18}. 
CCDDHNR (\citeyear{CCDDHNR18}) combine the use of Neyman orthogonality condition with cross fitting to provide a simple yet widely applicable framework that covers a large class of models under i.i.d. settings. 
The DML is also compatible with various types of machine learning based methods for nuisance parameter estimation. 

Driven by the need from empiricists, the literature on cluster robust inference has a long history in econometrics. 
For recent review of the literature, see, e.g., \cite{CM15} and \cite{MacKinnon2019}.
On the other hand, coping with cross-sectional dependence using a multiway cluster robust variance estimator is a relatively recent phenomenon. 
\cite{CGM11} first provide a multiway cluster robust variance estimator for linear regression models without imposing additional parametric assumptions on the intra-cluster correlation structure. 
This variance estimator has significantly reshaped the landscape of econometric practices in applied microeconomics in the past decade.\footnote{As of December 31, 2019, \cite{CGM11} has received over 2,500 citations. The majority of such citations came from applied economic papers.} 
 In contrast to the popularity among empirical researchers, theoretical justification of the validity of this type of procedures was lagging behind. 
The first rigorous treatment of asymptotic properties of multiway cluster robust estimators are established by \cite{Menzel17} using the Aldous-Hoover representation under the assumptions of separable exchangeability and dissociation. 
The asymptotic theory of \cite{Menzel17} covers both non-degenerate and degenerate cases. 
Focusing on non-degenerate situations, \cite{DDG18} further extend this approach to a general empirical process theory.\footnote{See also \cite{DDG19} for further generalization of the empirical process theory for dyadic data under joint exchangeability assumption.}
Using this asymptotic framework, \cite{MacKinnonNielsenWebb2019} study linear regression models under the non-degenerate case and examine the validity of several types of wild bootstrap procedures and the robustness of multiway cluster robust variance estimators under different cluster sampling settings.

Despite of the popularity of both machine learning and cluster robust inference among empirical researchers, relatively limited cluster robust inference results exist for machine learning based methods. 
Inference for machine learning based methods with one-way clustering is studied by \cite{BCHK16}, \cite{Kock2016}, \cite{KockTang2018}, \cite{SGCT18} and \cite{HansenLiao19} 
for different variations of regularized regression estimators and \cite{AtheyWager19} for random forests.
\cite{ChiangSasaki2019} investigate the performance of lasso and post-lasso in the partially linear model setting of \cite{BCH14} under multiway cluster sampling. 
To our best knowledge, there is no general machine learning based procedures with known validity under multiway cluster sampling environments.

\section{Overview}\label{sec:overview}
\subsection{Setup}\label{sec:setup}
Suppose that the researcher observes a sample $\left\{\left. W_{ij} \right\vert i \in \{1,...,N\}, j \in \{1,...,M\}\right\}$ of double-indexed observations of size $NM$. 
Let $P$ denote the probability law of $\{W_{ij}\}_{ij}$, and let $\Ep$ denote the expectation with respect to $P$. 
Let $\C= N \wedge M$ denote the sample size in the smaller dimension. 
We consider two-way clustering where each cell contains one observation for simplicity of notations, but results for higher cluster dimensions and random cluster sizes can be obtained at the expense of involved notations -- see Appendix \ref{sec:extension_to_general_multiway_clustering} for a general case.

The structural model is assumed to entail the moment restriction
\begin{align}
\Ep[\psi(W_{11};\theta_0,\eta_0)]=0
\label{eq:existance_condition}
\end{align} 
for some score $\psi$ that depends on a low-dimensional parameter vector $\theta \in \Theta \subset \Real^{d_\theta}$ and a nuisance parameter $\eta \in T$ for a convex subset $T$ of a normed linear space.
The nuisance parameter $\eta$ may be finite-, high-, or infinite-dimensional, and its true value is denoted by $\eta_0 \in T$.
In this setup, the true value of the low-dimensional target parameter, denoted by $\theta_0 \in \Theta$, is the object of interest.

Let $\tilde T=\{\eta - \eta_0 : \eta \in T\}$, and define the Gateaux derivative map $D_r: \tilde T \rightarrow \Real^{d_\theta}$ by
\begin{align*}
D_r[\eta-\eta_0]:=\partial_r \Big\{
\Ep[\psi(W_{11};\theta_0,\eta_0+r(\eta-\eta_0))]\Big\}
\end{align*}
for all $r\in[0,1)$.
Also denote its limit by
\begin{align*}
\partial_\eta\Ep\psi(W_{11};\theta_0,\eta_0)[\eta - \eta_0]:=D_0[\eta-\eta_0].
\end{align*} 
We say that the Neyman orthogonality condition holds at $(\theta_0,\eta_0)$ with respect to a nuisance realization set $\mathcal T_n \subset T$ if the score $\psi$ satisfies (\ref{eq:existance_condition}), the pathwise derivative $D_r[\eta-\eta_0]$ exists for all $r\in[0,1)$ and $\eta\in \mathcal T_n$, and the orthogonality equation
\begin{align}
\partial_\eta\Ep\psi(W_{11};\theta_0,\eta_0)[\eta - \eta_0]=0
\label{eq:Neyman_orthogonal_condition}
\end{align}
holds for all $\eta\in \mathcal T_n$. 
Furthermore, we also say that the $\lambda_n$ Neyman near-orthogonality condition holds at $(\theta_0,\eta_0)$ with respect to a nuisance realization set $\mathcal T_n\subset T$ if the score $\psi$ satisfies (\ref{eq:existance_condition}), the pathwise derivative $D_r[\eta-\eta_0]$ exists for all $r\in[0,1)$ and $\eta\in \mathcal T_n$, and the orthogonality equation
\begin{align}
\sup_{\eta \in \mathcal T_n}\Big\| \partial_\eta \Ep\psi(W;\theta_0,\eta_0)[\eta-\eta_0] \Big\|\le \lambda_n
\label{eq:Neyman_near_orthogonal_condition}
\end{align}
holds for all $\eta\in \mathcal T_n$
for some positive sequence $\{\lambda_n\}_n$ such that $\lambda_n=o(\C^{-1/2})$.

Throughout, we will consider structural models satisfying the moment restriction (\ref{eq:existance_condition}) and either form of the Neyman orthogonality conditions, (\ref{eq:Neyman_orthogonal_condition}) or (\ref{eq:Neyman_near_orthogonal_condition}).
Consider linear Neyman orthogonal scores $\psi$ of the form
\begin{align}
\psi(w;\theta,\eta)=\psi^a(w;\eta)\theta +\psi^b(w;\eta), \text{ for all $w\in \supp(W)$, $\theta\in\Theta$, $\eta\in T$. } \label{eq:linear_score}
\end{align}
A generalization to nonlinear score follows from linearization with Gateaux differentiability as in Section 3.3 of CCDDHNR (\citeyear{CCDDHNR18}).
We focus on linear scores as they cover a wide range of applications.

\subsection{The Multiway Double/Debiased Machine Learning}\label{sec:multiway_dml}
For the class of models introduced in Section \ref{sec:setup}, we propose a novel $K^2$-fold multiway cross fitting procedure for estimation of $\theta_0$.
For any $r \in \mathbb N$, we use the notation $[r]=\{1,...,r\}$. 
With a fixed positive integer $K$, randomly partition $[N]$ into $K$ parts $\{I_1,...,I_K\}$ and $[M]$ into $K$ parts $\{J_1,...,J_K\}$. 
For each $(k,\ell) \in [K]^2$, obtain an estimate
$$\hat \eta_{k\ell}=\hat \eta\left((W_{ij})_{(i,j)\in ([N]\setminus I_k )\times ([M]\setminus J_\ell)}\right)$$ 
of the nuisance parameter $\eta$ by some machine learning method (e.g., lasso, post-lasso, elastic nets, ridge, deep neural networks, and boosted trees) using only the subsample of those observations with multiway indices $(i,j)$ in $([N]\setminus I_k ) \times ([M]\setminus J_\ell)$.
In turn, we define $\tilde \theta$, the multiway double/debiased machine learning (multiway DML) estimator for $\theta_0$, as the solution to 
\begin{align}
\frac{1}{K^2}\sumkl \Enkl[\psi(W;\tilde \theta,\hat \eta_{k\ell})] =0,\label{eq:MDML}
\end{align}
where $\Enkl [f(W)] = \frac{1}{|I_k||J_\ell|}\sumIJ f(W_{ij})$ denotes the subsample empirical expectation using only the those observations with multiway indices $(i,j)$ in $I_k \times J_\ell$.

\begin{figure}
\caption{An illustration of $2^2$-fold cross fitting.}\label{fig:cross_fitting}
\tikzstyle{my help lines}=[gray,
thick,dashed]
\begin{multicols}{4}
\qquad\\
\begin{tikzpicture}
\draw(4,0)  grid (3,1)node[above,black]{Nuisance};
\draw (3,1)  grid (2,2);
\draw (3,0)  grid (2,1);
\draw (3,2)  grid (4,1);
\draw (0,4) grid (1,3)node[above,black]{Score};
\draw (1,4) grid (2,3);
\draw (0,3) grid (1,2);
\draw (1,3) grid (2,2);
\draw[style=my help lines] (2,0) grid (0,2);
\draw[style=my help lines] (4,2) grid (2,4);
\end{tikzpicture}
\qquad\\

\begin{tikzpicture}

\draw(4,2)  grid (3,3)node[above,black]{Nuisance};
\draw(3,3)  grid (2,4);
\draw(4,3) grid (3,4);
\draw(3,2) grid (2,3);

\draw(2,0)  grid (1,1)node[above,black]{Score};
\draw(1,1)  grid (0,2);
\draw(2,1) grid(0,2);
\draw(1,0) grid(0,1);
\draw[style=my help lines] (4,0) grid (2,2);
\draw[style=my help lines] (2,2) grid (0,4);
\end{tikzpicture}
\qquad\\

\begin{tikzpicture}

\draw(4,2) [darkgray] grid (3,3)node[above,black]{Score};
\draw(3,3) [darkgray] grid (2,4);
\draw(4,3) [darkgray] grid (3,4);
\draw(3,2) [darkgray] grid (2,3);

\draw(2,0) [darkgray] grid (1,1)node[above,black]{Nuisance};
\draw(1,1) [darkgray] grid (0,2);
\draw(2,1) [darkgray] grid(0,2);
\draw(1,0) [darkgray] grid(0,1);
\draw[style=my help lines] (4,0) grid (2,2);
\draw[style=my help lines] (2,2) grid (0,4);
\end{tikzpicture}
\qquad\\

\begin{tikzpicture}

\draw (4,0) [darkgray] grid (3,1)node[above,black]{Score};
\draw (3,1) [darkgray] grid (2,2);
\draw (3,0) [darkgray] grid (2,1);
\draw (3,2) [darkgray] grid (4,1);
\draw (0,4) [darkgray] grid (1,3)node[above,black]{Nuisance};
\draw (1,4) [darkgray] grid (2,3);
\draw (0,3) [darkgray] grid (1,2);
\draw (1,3) [darkgray] grid (2,2);
\draw[style=my help lines] (2,0) grid (0,2);
\draw[style=my help lines] (4,2) grid (2,4);
\end{tikzpicture}
\end{multicols}
\end{figure}
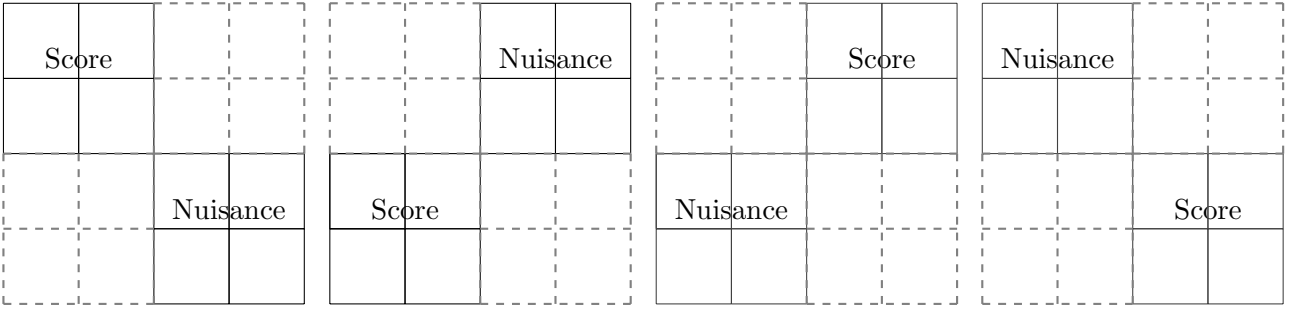

We call this procedure the $K^2$-fold multiway cross fitting.
Note that, for each $(k,\ell)\in [K]^2$, the nuisance parameter estimate $\hat\eta_{k\ell}$ is computed using the subsample of those observations with multiway indices $(i,j) \in ([N]\setminus I_k ) \times ([M]\setminus J_\ell)$, and in turn the score term $\Enkl[\psi(W; \cdot,\hat\eta_{k\ell})]$ is computed using the subsample of those observations with multiway indices $(i,j) \in I_k \times J_\ell$.
This two-step computation is repeated $K^2$ times for every partitioning pair $(k,\ell)\in [K]^2$. 
Figure \ref{fig:cross_fitting} illustrates this $K^2$-fold cross fitting for the case of $K=2$ and $N=M=4$, where the cross fitting repeats for $K^2 (= 2^2 = 4)$ times.

\begin{remark}
This estimator is a multiway-counterpart of DML2 in CCDDHNR (\citeyear{CCDDHNR18}). 
It is also possible to consider the multiway-counterpart of their DML1. 
With this said, we focus on this current estimator following their simulation finding that DML2 outperforms their DML1 in most situation settings due to the stability of the score function.  
\end{remark}
\begin{remark}[Higher Cluster Dimensions]\label{remark:higher_cluster_dim}
When we have $\alpha$-way clustering for an integer $\alpha>2$, the above algorithm can be easily generalized into a $K^\alpha$-fold multiway DML estimator. See Appendix \ref{sec:extension_to_general_multiway_clustering} for a generalization.
\end{remark}

We propose to estimate the asymptotic variance of $\sqrt{\C}(\tilde\theta-\theta_0)$ by
\begin{align}
\hat\sigma^2
=&
\hat J^{-1} 
 \hat\Gamma
(\hat J^{-1})',
\label{eq:variance_estimator}
\end{align}
where $\hat \Gamma$ and $\hat J$ are given by
\begin{align}
\hat \Gamma
=&
\frac{1}{K^2}\sumkl \left\{
\frac{|I|\wedge|J|}{(|I||J|)^2}\sum_{i\in I_k}\sum_{j,j'\in J_\ell} \psi(W_{ij};\tilde \theta,\hat\eta_{k\ell}) \psi(W_{ij'};\tilde \theta,\hat\eta_{k\ell})'
\right.
\nonumber\\
& \qquad\qquad \ \
+
\left.
\frac{|I|\wedge|J|}{(|I||J|)^2}\sum_{i,i'\in I_k}\sum_{j\in J_\ell} \psi(W_{ij};\tilde \theta,\hat\eta_{k\ell}) \psi(W_{i'j};\tilde \theta,\hat\eta_{k\ell}) '
\right\}
\qquad\text{and}
\nonumber\\
\hat J 
=&
\frac{1}{K^2}\sumkl\Enkl[\psi^a(W;\hat \eta_{k\ell})],\nonumber
\end{align}
accounting for multiway cluster dependence.
For a $d_\theta$-dimensional vector $r$, the $(1-a)$ confidence interval for the linear functional  $r'\theta_0$ can be constructed by
\begin{align*}
\text{CI}_a:=[r'\tilde \theta\pm \Phi^{-1}(1-a/2)\sqrt{r'\hat \sigma^2 r/\C}].
\end{align*}

\subsection{Example: Partially Linear IV Model with Multiway Cluster Sample}\label{sec:example_partially_linear}
For an illustration, consider as a concrete example the partially linear IV model (cf. Okui, Small, Tan and Robins, \citeyear{OkuiSmallTanRobins2012} ; CCDDHNR, \citeyear{CCDDHNR18}, Section 4.2) adapted to the multiway cluster sample data:
\begin{align}
Y_{ij}=&D_{ij}\theta_0 + g_0(X_{ij})+\epsilon_{ij},\qquad\Ep[\epsilon_{ij}|X_{ij},Z_{ij}]=0,
\label{eq:example:reduced_form}
\\
Z_{ij}=&m_0(X_{ij})+v_{ij},\quad\:\qquad\qquad\Ep[v_{ij}|X_{ij}]=0.
\label{eq:example:projection}
\end{align}
A researcher observes the random variables $Y_{ij}$, $D_{ij}$, $X_{ij}$, and $Z_{ij}$, which are typically interpreted as the outcome, endogenous regressor, exogenous regressors, and instrumental variable, respectively.
The low-dimensional parameter vector $\theta_0$ is an object of interest.

A Neyman orthogonal score $\psi$ for such model is given by
\begin{align}
\psi(w;\theta,\eta)=(y-g_1(x)-\theta(d - g_2(x)))(z-m(x))
\label{eq:IV_Neymand_orthogonal_moment}
\end{align}
as in \cite{OkuiSmallTanRobins2012} and CCDDHNR (\citeyear{CCDDHNR18}), 
where $w=(y,d,x,z)$, $\eta=(g_1,g_2,m)$ and $g_1$, $g_2$, $m\in L^2(P)$. 
It is straightforward to verify that this score satisfies both the moment restriction (\ref{eq:existance_condition}), $\Ep[\psi(W_{11};\theta_0,\eta_0)]=0$, and the Neyman orthogonality condition (\ref{eq:Neyman_orthogonal_condition}), $\partial_\eta \Ep \psi(W_{11};\theta_0,\eta_0)[\eta - \eta_0]=0$ for all $\eta \in \mathcal{T}_n$ at $\eta_0=(g_{10},g_{20},m_0)$, where $g_{10}(X)=\Ep[Y|X]$, $g_{20}(X)=\Ep[D|X]$, and $m_0(X)=\Ep[Z|X]$.

The following algorithm is our proposed multiway DML procedure introduced in Section \ref{sec:multiway_dml}, specifically applied to this partially linear IV model.
\begin{algorithm}[$K^2$-fold Multiway DML for Partially Linear IV Model with Lasso]\label{algorithm:partial_linear_iv}
${}$
\begin{enumerate}
\item Randomly partition $[N]$ into $K$ parts $\{I_1,...,I_K\}$ and $[M]$ into $K$ parts $\{J_1,...,J_K\}$.   
\item For each $(k,\ell)\in[K]^2$:
\begin{enumerate}
\item Run a lasso of $Y$ on $X$ to obtain $\hat g_{1,k\ell}(x)=x'\hat\beta_{k\ell}$ using observations from $I_k^c\times J_\ell^c$.
\item Run a lasso of $D$ on $X$ to obtain $\hat g_{2,k\ell}(x)=x'\hat\gamma_{k\ell}$ using observations from $I_k^c\times J_\ell^c$.
\item Run a lasso of $Z$ on $X$ to obtain $\hat m_{k\ell}(x)=x'\hat \xi_{k\ell}$ using observations from $I_k^c\times J_\ell^c$.
\end{enumerate}
\item Solve the equation
\begin{align*}
\frac{1}{K^2}\sum_{(k,\ell) \in [K]^2}\Enkl[(Y_{ij}-X_{ij}'\hat\beta_{k\ell}-\theta(D_{ij} - X_{ij}'\hat \gamma_{k\ell}))(Z_{ij}-X_{ij}'\hat\xi_{k\ell})]=0
\end{align*}
for $\theta$ to obtain the multiway DML estimate $\tilde\theta$.
\item Let $\hat \varepsilon_{ij}=Y_{ij}- X_{ij}' \hat\beta_{k \ell} - \tilde\theta (D_{ij}-X_{ij}'\hat\gamma_{k\ell})$, $\hat u_{ij} = D_{ij} - X_{ij}'\hat\gamma_{k\ell}$, and $\hat v_{ij}=Z_{ij}-X_{ij}'\hat \xi_{k \ell}$ for each $(i,j) \in I_k \times J_\ell$ for each $(k,\ell) \in [K]^2$, and let the multiway DML asymptotic variance estimator be given by
\begin{align*}
\hat \sigma^2=&\hat J^{-1} \frac{1}{K^2}\sum_{k=1}^K \sum_{\ell=1}^K\Big\{
\frac{|I|\wedge |J|}{(|I||J|)^2}\sum_{i\in I_k}\sum_{j,j'\in J_\ell}
\hat \varepsilon_{ij}\hat v_{ij} \hat v_{ij'} \hat \varepsilon_{ij'}
+
\frac{|I|\wedge |J|}{(|I||J|)^2}\sum_{i,i'\in I_k}\sum_{j\in J_\ell} \hat \varepsilon_{ij}\hat v_{ij} \hat v_{i'j} \hat \varepsilon_{i'j}
 \Big\}(\hat J^{-1})',
\end{align*}
where
\begin{align*}
\hat J=&-\frac{1}{K^2}\sum_{k=1}^K\sum_{\ell=1}^K\Enkl[\hat u_{ij}\hat v_{ij} ].
\end{align*}
\item Report the estimate $\tilde\theta$, its standard error $\sqrt{\hat\sigma^2/\C}$, and/or the $(1-a)$ confidence interval
\begin{align*}
\text{CI}_a:=\left[\tilde \theta\pm \Phi^{-1}(1-a/2)\sqrt{\hat \sigma^2 /\C}\right].
\end{align*}
\end{enumerate}
\end{algorithm}

For the sake of concreteness, we present this algorithm specifically based on lasso (in the three sub-steps under step 2), but another machine learning method (e.g., post-lasso, elastic nets, ridge, deep neural networks, and boosted trees) may be substituted for lasso.

\begin{example}[Demand Analysis]\label{ex:demand_analysis}
Consider the model of \citet{Berry94}
in which consumer $c$ derives the utility
\begin{align*}
\delta_{ij} + X_{ij}\alpha_c + \varepsilon_{cij}
\end{align*}
from choosing product $i$ in market $j$,
where $\varepsilon_{cij}$ independently follows the Type I Extreme Value distribution, $\alpha_c$ is a random coefficient, and the mean utility $\delta_{ij}$ takes the linear-index form
\begin{align*}
\delta_{ij} = D_{ij}\theta_0 + \epsilon_{ij}.
\end{align*}
In this framework, \citet[][Equation (9)]{LuShiTao19} derive the partial-linear equation
\begin{align*}
Y_{ij} = D_{ij}\theta_0 + g_0(X_{ij}) + \epsilon_{ij}
\end{align*}
for estimation of $\theta_0$, where $Y_{ij} = \log( S_{ij} ) - \log( S_{0j} )$ denotes the observed log share of product $i$ relative to the log of the outside share.
Since $D_{ij}$ usually consists of the endogenous price of product $i$ in market $j$, researchers often use instruments $Z_{ij}$ such that $\Ep[\epsilon_{ij}|X_{ij},Z_{ij}]=0$.
This yields the reduced-form equation (\ref{eq:example:reduced_form}), together with the innocuous nonparametric projection equation (\ref{eq:example:projection}).
Since the random vector $W_{ij} = (Y_{ij},D_{ij},X_{ij},Z_{ij})$ is double-indexed by product $i$ and market $j$, the sample naturally entails two-way dependence. 
Specifically, for each product $i$, $\{W_{ij}\}_{j=1}^M$ is likely dependent through a supply shock by the producer of product $i$.
Similarly, for each market $j$, $\{W_{ij}\}_{i=1}^N$ is likely dependent through a demand shock in market $j$.
As such, instead of using standard errors based on i.i.d. sampling, we recommend that a researcher uses the two-way cluster-robust standard error based on Algorithm \ref{algorithm:partial_linear_iv}.
$\triangle$
\end{example}

\section{Theory of the Multiway DML}\label{sec:Theory of the Multiway DML}

In this section, we present formal theories to guarantee that the multiway DML method proposed in Section \ref{sec:overview} works.
We first fix some notations for convenience. 
The two-way sample sizes $(N,M) \in \mathbb{N}^2$ will be index by a single index $n \in \mathbb{N}$ as $(N,M) = (N(n),M(n))$ where $M(n)$ and $N(n)$ are non-decreasing in $n$ and $M(n)N(n)$ is increasing in $n$.
With this said, we will suppress the index notation and write $(N,M)$ for simplicity.
Let $\{\mathcal P_n\}_n$ be a sequence of sets of probability laws of $\{W_{ij}\}_{ij}$ -- note that we allow for increasing dimensionality of $W_{ij}$ in the sample size $n$. 
Let $P=P_{n}\in \mathcal P_n$ denote the law with respect to sample size $(N,M)$.
Throughout, we assume that this random vector $W_{ij}$ is Borel measurable. 
Recall the notations $\C =N\wedge M$, $\mu_N=\C/N$, and $\mu_M=\C/M$, and suppose that $\mu_N\to \bar \mu_N$, $\mu_M\to \bar \mu_M$.
We write $a \lesssim b$ to mean $a \leq cb$ for some $c > 0$ that does not depend on $n$.
We also write $a \lesssim_P b$ to mean $a = O_P(b)$.
For any finite dimensional vector $v$, $\|v\|$ denotes the $\ell_2$ or Euclidean norm of $v$.
For any matrix $A$, $\|A\|$ denotes the induced $\ell_2$-norm of the matrix. For any set $B$,  $|B|$ denotes the cardinality of the set.

We state the following assumption on multiway clustered sampling.
\begin{assumption}[Sampling]\label{a:sampling}
Suppose $\C \to \infty $. 
The following conditions hold for each $n$.
\begin{enumerate}[(i)]
\item $(W_{ij})_{(i,j)\in \mathbbm N^2}$ is an infinite sequence of separately exchangeable $p$-dimensional random vectors. 
That is, for any permutations $\pi_1$ and $\pi_2$ of $\mathbbm N$, we have
\begin{align*}
(W_{ij})_{(i,j)\in \mathbbm N^2}\overset{d}{=} (W_{\pi_1(i)\pi_2(j)})_{(i,j)\in \mathbbm N^2}.
\end{align*}
\item $(W_{ij})_{(i,j)\in \mathbbm N^2}$ is dissociated. 
That is, for any $(c_1,c_2)\in \mathbbm N^2$, 
$
(W_{ij})_{i \in [c_1], j \in [c_2]}
$
is independent of 
$
(W_{ij})_{i \in [c_1]^c, j \in [c_2]^c}.
$
\item For each $n$, an econometrician observes $(W_{ij})_{i\in[N],j\in[M]}$.
\end{enumerate}
\end{assumption}
Recall that we focus on the linear Neyman orthogonal score of the form
\begin{align*}
\psi(w;\theta,\eta)=\psi^a(w;\eta)\theta +\psi^b(w;\eta), \text{ for all $w\in \supp(W)$, $\theta\in\Theta$, $\eta\in T$. }
\end{align*}
Let $c_0>0$, $c_1>0$, $s>0$, $q\ge 4$ 
be some finite constants with $c_0\le c_1$. 
Let $\{\delta_n\}_{n\ge 1}$ (estimation errors) and $\{\Delta_n\}_{n\ge 1}$ (probability bounds) be sequences of positive constants that converge to zero such that $\delta_n \ge \C^{-1/2}$. 
Let $K\ge 2$ be a fixed integer. 
Let $W_{00}$ denote a copy of $W_{11}$ that is independent from the data and the random set $\mathcal T_n$ of nuisance realization. 
With these notations, we consider the following assumptions.
\begin{assumption}[Linear Neyman Orthogonal Score]\label{a:linear_orthogonal_score}
For $\C\ge 3$ and $P\in \mathcal P_n$, the following conditions hold.
\begin{enumerate}[(i)]
\item The true parameter value $\theta_0$ satisfies (\ref{eq:existance_condition}).
\item $\psi$ is linear in the sense that it satisfies (\ref{eq:linear_score}).
\item The map $\eta \mapsto \Ep[\psi(W_{00};\theta,\eta)]$ is twice continuously Gateaux differentiable on $T$.
\item $\psi$ satisfies either the Neyman orthogonality condition (\ref{eq:Neyman_orthogonal_condition}) or more generally 
the Neyman $\lambda_n$ near orthogonality condition at $(\theta_0,\eta_0)$ with respect to a nuisance realization set $\mathcal T_n\subset T$ as
\begin{align*}
\lambda_n:=\sup_{\eta \in \mathcal T_n}\Big\| \partial_\eta \Ep\psi(W_{00};\theta_0,\eta_0)[\eta-\eta_0] \Big\|\le \delta_n \C^{-1/2}.
\end{align*}
\item The identification condition holds as the singular values of the matrix $J_0:=\Ep[\psi^a(W_{11};\eta_0)]$ are between $c_0$ and $c_1$.
\end{enumerate}
\end{assumption}
\begin{assumption}[Score Regularity and Nuisance Parameter Estimators]\label{a:regularity_nuisance_parameters}
For all $\C\ge 3$ and $P\in \mathcal P_n$, the following conditions hold.
\begin{enumerate}[(i)]
\item Given random subsets $I\subset [N]$ and $J\subset [M]$ such that $|I|\times |J|=\lfloor NM/K^2\rfloor$, the nuisance parameter estimator $\hat \eta=\hat\eta((W_{ij})_{(i,j)\in I^c\times J^c}) $, where the complements are taken with respect to $[N]$ and $[M]$, respectively, belongs to the realization set $\mathcal T_n$ with probability at least $1-\Delta_n$, where $\mathcal T_n$ contains $\eta_0 $.
\item The following moment conditions hold:
\begin{align*}
m_n:=& \sup_{\eta\in \T_n}(\Ep[\|\psi(W_{00};\theta_0,\eta)\|^q])^{1/q} \le c_1,\\
m_n':=& \sup_{\eta\in \T_n}(\Ep[\|\psi^a(W_{00};\eta)\|^q])^{1/q} \le c_1.
\end{align*}
\item The following conditions on the rates $r_n$, $r_n'$ and $\lambda_n'$ hold:
\begin{align*}
r_n:=& \sup_{\eta\in \T_n}
\|\Ep[\psi^a(W_{00};\eta)]-\Ep[\psi^a(W_{00};\eta_0)]\|\le \delta_n,\\
r_n':=& \sup_{\eta\in \T_n}
(\|\Ep[\psi(W_{00};\theta_0,\eta)]-\Ep[\psi(W_{00};\theta_0,\eta_0)]\|^2)^{1/2}\le \delta_n,\\
\lambda_n'= & \sup_{r\in (0,1),\eta\in \T_n}\|\partial^2_r \Ep[\psi (W_{00};\theta_0,\eta_0+r(\eta-\eta_0)) ] \|\le \delta_n/\sqrt{\C}.
\end{align*}
\item All eigenvalues of the matrix 
\begin{align*}
\Gamma:=\bar\mu_N \Gamma_N + \bar\mu_M \Gamma_M=\bar\mu_N\Ep [\psi(W_{11};\theta_0,\eta_0)\psi(W_{12};\theta_0,\eta_0)'] 
+ \bar\mu_M\Ep[\psi(W_{11};\theta_0,\eta_0)\psi(W_{21};\theta_0,\eta_0)'].   
\end{align*}
are bounded from below by $c_0$.
\end{enumerate}
\end{assumption}

\begin{remark}[Discussion of the Assumptions]
Assumption \ref{a:sampling} is similar to those of the preceding work on multiway cluster robust inference \citep[cf.][]{Menzel17,DDG18,ChiangSasaki2019}.
\citet{Menzel17} does not invoke the dissociation, and follows an alternative approach to inference.
The other papers assume both the separate exchangeability and dissociation, and conduct unconditional inference as in this paper.
See \citet[][Corollary 7.23 and Lemma 7.35]{Kallenberg2006} for representations with and without the dissociation under the separate exchangeability.
Assumption \ref{a:linear_orthogonal_score}
is closely related to Assumptions 3.1 of CCDDHNR (\citeyear{CCDDHNR18}). It requires the score to be 
Neyman near orthogonal -- see their Section 2.2.1 for the procedure of orthogonalizing a non-orthogonal score. 
It also imposes some mild smoothness and identification conditions.
Assumption \ref{a:regularity_nuisance_parameters} corresponds to Assumption 3.2 of CCDDHNR (\citeyear{CCDDHNR18}). It imposes some high level conditions on the quality of the nuisance parameter estimator as well as the non-degeneracy of the asymptotic variance. This rules out the degenerate cases such as Example 1.6 of \cite{Menzel17}.
\end{remark}

\begin{remark}[Partial Distributions]
Assumptions \ref{a:linear_orthogonal_score} and \ref{a:regularity_nuisance_parameters} state conditions based on $W_{00}$, differently from CCDDHNR (\citeyear{CCDDHNR18}), because of our need to deal with dependent observations in cross fitting in our multiway DML framework.
\end{remark}

The following result presents the main theorem of this paper, establishing the linear representation and asymptotic normality of the multiway DML estimator.  
It corresponds to Theorem 3.1 of CCDDHNR (\citeyear{CCDDHNR18}), and is an extension of it to the case of multiway cluster sampling.

\begin{theorem}[Main Result]\label{theorem:main_result_linear}
Suppose that Assumptions \ref{a:sampling}, \ref{a:linear_orthogonal_score} and \ref{a:regularity_nuisance_parameters} are satisfied.
If $\delta_n\ge \C^{-1/2}$ for all $\C\ge 1$, then
\begin{align*}
\sqrt{\C}\sigma^{-1}(\tilde \theta - \theta_0)=\frac{\sqrt{\C}}{NM}\sumij
\bar \psi(W_{ij})+\Op(\rho_n)\leadsto N(0,I_{d_\theta})
\end{align*}
holds uniformly over $P\in\mathcal P_n$, where the size of the remainder terms follows
\begin{align*}
\rho_n :=\C^{-1/2} + r_n +r_n' + \C^{1/2} \lambda_n + \C^{1/2} \lambda_n'\lesssim \delta_n,
\end{align*}
the influence function takes the form $\bar \psi(\cdot):=-\sigma^{-1}J_0^{-1} \psi(\cdot;\theta_0,\eta_0)$, 
and the asymptotic variance is given by
\begin{align}
\sigma^2:=J_0^{-1}\Gamma (J_0^{-1})'. \label{eq:population_variance}
\end{align}
\end{theorem}

As is commonly the case in practice, we need to estimate the unknown asymptotic variance.
The following theorem shows the validity of our proposed multiway DML variance estimator.

\begin{theorem}[Variance Estimator]\label{theorem:variance_estimator_linear}
Under the assumptions required by Theorem \ref{theorem:main_result_linear}, we have
\begin{align*}
\hat \sigma^2=\sigma^2 +\Op(\rho_n).
\end{align*}
Furthermore, the statement of Theorem \ref{theorem:main_result_linear} holds true with $\hat \sigma^2$ in place of $\sigma^2$.
\end{theorem}

Theorems \ref{theorem:main_result_linear} and \ref{theorem:variance_estimator_linear} can be used for constructing confidence intervals.

\begin{corollary}\label{corollary:inference_t-test}
Suppose that all the Assumptions required by Theorem \ref{theorem:main_result_linear} are satisfied.
Let $r$ be a $d_\theta$-dimensional vector. 
The $(1-a)$ confidence interval of $r'\theta_0$ given by
\begin{align*}
\text{CI}_a:=[r'\tilde \theta\pm \Phi^{-1}(1-a/2)\sqrt{r'\hat \sigma^2 r/\C}]
\end{align*}
satisfies
\begin{align*}
\sup_{P\in\mathcal P_n}|P_P(\theta_0 \in \text{CI}_a)-(1-a)|\to 0.
\end{align*}
\end{corollary}

As in Section 3.4 of CCDDHNR (\citeyear{CCDDHNR18}), we can also repeatedly compute multiway DML estimates and variance estimates $S$-times for some fixed $S\in \mathbbm N$ and consider the average or median of the estimates as the new estimate. 
This does not have an asymptotic impact, yet it can reduce the impact of a random sample splitting on the estimate.

\section{Simulation Studies}\label{sec:simulation_studies}
\subsection{Simulation Setup}
Consider the partially linear IV model introduced in Section \ref{sec:example_partially_linear}.
We specifically focus on the following high-dimensional linear representations
\begin{align*}
Y_{ij} =& D_{ij}\theta_0 + X_{ij}'\zeta_0 + \epsilon_{ij}
\\
D_{ij} =& Z_{ij}\pi_{10} + X_{ij}'\pi_{20} + \upsilon_{ij},
\\
Z_{ij} =& X_{ij}'\xi_0 + V_{ij},
\end{align*}
where the parameter values are set to $\theta_0 = \pi_{10} = 1.0$ and $\zeta_0 = \pi_{20} = \xi_0 = (0.5,.0.5^2,\cdots,0.5^{\text{dim}(X)})'$ for some large $\text{dim}(X)$.
The primitive random vector $(X_{ij}',\epsilon_{ij},\upsilon_{ij},V_{ij})'$ is constructed by
\begin{align*}
X_{ij} &= (1-\omega_1^X - \omega_2^X) \alpha_{ij}^X + \omega_1^X \alpha_i^X + \omega_2^X \alpha_j^X,
\\
\epsilon_{ij} &= (1-\omega_1^\epsilon - \omega_2^\epsilon) \alpha_{ij}^\epsilon + \omega_1^\epsilon \alpha_i^\epsilon + \omega_2^\epsilon \alpha_j^\epsilon,
\\
\upsilon_{ij} &= (1-\omega_1^\upsilon - \omega_2^\upsilon) \alpha_{ij}^\upsilon + \omega_1^\upsilon \alpha_i^\upsilon + \omega_2^\upsilon \alpha_j^\upsilon,
\qquad\text{and}\\
V_{ij} &= (1-\omega_1^V - \omega_2^V) \alpha_{ij}^V + \omega_1^V \alpha_i^V + \omega_2^V \alpha_j^V
\end{align*}
with two-way clustering weights $(\omega_1^X,\omega_2^X)$, $(\omega_1^\epsilon,\omega_2^\epsilon)$, $(\omega_1^\upsilon,\omega_2^\upsilon)$, and $(\omega_1^V,\omega_2^V)$, where 
$\alpha_{ij}^X$, $\alpha_{i}^X$, and $\alpha_{j}^X$ are independently generated according to
\begin{align*}
\alpha_{ij}^X, \alpha_{i}^X, \alpha_{j}^X
\sim N
\left(0, \left(\begin{array}{ccccc}
s_X^0 & s_X^1 & \cdots & s_X^{\text{dim}(X)-2} & s_X^{\text{dim}(X)-1} \\
s_X^1 & s_X^0 & \cdots & s_X^{\text{dim}(X)-3} & s_X^{\text{dim}(X)-2} \\
\vdots & \vdots & \ddots & \vdots & \vdots \\
s_X^{\text{dim}(X)-2} & s_X^{\text{dim}(X)-3} & \cdots & s_X^0 & s_X^1  \\
s_X^{\text{dim}(X)-1} & s_X^{\text{dim}(X)-2} & \cdots & s_X^1 & s_X^0
\end{array}\right)\right),
\end{align*}
$(\alpha_{ij}^\epsilon,\alpha_{ij}^\upsilon)'$, $(\alpha_{i}^\epsilon,\alpha_{i}^\upsilon)'$, and $(\alpha_{j}^\epsilon,\alpha_{j}^\upsilon)'$ are independently generated according to
\begin{align*}
\left(\begin{array}{c}\alpha_{ij}^\epsilon \\ \alpha_{ij}^\upsilon\end{array}\right),
\left(\begin{array}{c}\alpha_{i}^\epsilon \\ \alpha_{i}^\upsilon\end{array}\right),
\left(\begin{array}{c}\alpha_{j}^\epsilon \\ \alpha_{j}^\upsilon\end{array}\right)
\sim N
\left(0, \left(\begin{array}{cc}
1 & s_{\epsilon\upsilon} \\
s_{\epsilon\upsilon} & 1
\end{array}\right)\right),
\end{align*}
and $\alpha_{ij}^V$, $\alpha_{i}^V$, and $\alpha_{j}^V$ are independently generated according to
\begin{align*}
\alpha_{ij}^V, \alpha_{i}^V, \alpha_{j}^V
\sim
N(0,1).
\end{align*}

The weights $(\omega_1^X,\omega_2^X)$, $(\omega_1^\epsilon,\omega_2^\epsilon)$, $(\omega_1^\upsilon,\omega_2^\upsilon)$, and $(\omega_1^V,\omega_2^V)$ specify the extent of dependence in two-way clustering in $X_{ij}$, $\epsilon_{ij}$, $\upsilon_{ij}$, and $V_{ij}$, respepctively.
The parameter $s_X$ specifies the extent of collinearity among the high-dimensional regressors $X_{ij}$.
The parameter $s_{\epsilon\upsilon}$ specifies the extent of endogeneity.
We set the values of these parameters to $(\omega_1^X,\omega_2^X) = (\omega_1^\epsilon,\omega_2^\epsilon) = (\omega_1^\upsilon,\omega_2^\upsilon) = (\omega_1^V,\omega_2^V) = (0.25, 0.25)$ and $s_X = s_{\epsilon\upsilon} = 0.25$.

\subsection{Results}
Monte Carlo simulations are conducted with 2,500 iterations for each set.
Table \ref{tab:simulation_results} reports simulation results.
The first four columns in the table indicate the data generating process ($N$, $M$, $\C$, and dim$(X)$).
The next column indicates the integer $K$ for our $K^2$-fold cross fitting method. 
We use $K=2$ and $3$ in the simulations for the displayed results, since $2^2 (\approx 5)$ and $3^2 (\approx 10)$ are close to the common numbers of folds used in cross fitting in practice.
The next column indicates the machine learning method for estimation of $\hat\eta_{k\ell}$.
We use the ridge, elastic net, and lasso.
The last four columns of the table report Monte Carlo simulation statistics, including the bias (Bias), standard deviation (SD), root mean square error (RMSE), and coverage frequency for the nominal probability of 95\% (Cover).

For each covariate dimension $\text{dim}(X) \in \{100,200\}$, for each choice $K \in \{2,3\}$ for the number $K^2$ of multiway cross fitting, and for each of the three machine learning methods, we observe the following patterns as the effective sample size $\C=N \wedge M$ increases: 1) the bias tends to zero; 2) the standard deviation decreases approximately at the $\sqrt{\C}$ rate; and 3) the coverage frequency converges to the nominal probability.
These results confirm the theoretical properties of the proposed method.
We ran several other sets of simulations besides those displayed in the table, and this pattern remains the same across different sets.

Comparing the results across the three machine learning methods, we observe that the ridge entails larger bias and smaller variance relative to the elastic net and lasso in finite sample.
This makes the coverage frequency of the ridge less accurate compared with the elastic net and lasso.
This result is perhaps specific to the data generating process used for our simulations.
On one hand, the choice $K=3$ (i.e., $9$-fold) of the multiway cross fitting contributes to mitigating the large bias of the ridge relative to the choice $K=2$, and hence $K=3$ produces more preferred results for the ridge.
On the other hand, the choice $K=2$ tends to yield preferred results in terms of coverage accuracy for the elastic net and lasso. 
In light of these results, we recommend the elastic net or lasso along with the use of $2^2$- fold (i.e., $4$-fold) cross fitting.
This number of folds in cross fitting is in fact similar to that recommended by CCDDHNR (\citeyear{CCDDHNR18}) for i.i.d. sampling -- see their Remark 3.1 where they recommend 4- or 5-fold cross fitting.

\section{Empirical Illustration: Demand Analysis with Market Share Data}\label{sec:empirical_illustration}

Let us revisit the demand model of Example \ref{ex:demand_analysis} in Section \ref{sec:example_partially_linear}.
Recall that, for the consumer demand model of \citet{Berry94} introduced in Example \ref{ex:demand_analysis}, \citet[][Equation (9)]{LuShiTao19} derive the partial-linear equation
\begin{align}\label{eq:partial_lienar_demand}
Y_{ij} = D_{ij}\theta_0 + g_0(X_{ij}) + \epsilon_{ij}
\end{align}
for estimation of $\theta_0$, where $Y_{ij} = \log( S_{ij} ) - \log( S_{0j} )$ denotes the observed log share of product $i$ relative to the log of the outside share in market $j$, $D_{ij}$ denotes the log price of product $i$ in market $j$, and $X_{ij}$ denotes a vector of observed attributes of product $i$ in market $j$.
To deal with the likely endogeneity of $D_{ij}$, researchers often use instruments $Z_{ij}$ such that $\Ep[\epsilon_{ij}|X_{ij},Z_{ij}]=0$.
Such instruments often consist of observed attributes of other products in the market.

The implied equation (\ref{eq:partial_lienar_demand}) together with this mean independence assumption yields the reduced-form model (\ref{eq:example:reduced_form}). 
Furthermore, we write the innocuous nonparametric projection equation (\ref{eq:example:projection}).
Therefore, we apply Algorithm \ref{algorithm:partial_linear_iv} in Section \ref{sec:example_partially_linear} for the two-way cluster robust DML estimation of $\theta_0$ with a robust standard error.

We present an application of the proposed algorithm to the U.S. automobile data of \citet{BLP95}.
The sample consists of unbalanced two-way clustered observations with $N=557$ models of automobiles and $M=20$ markets.
The observed attributes $X_{ij}$ consist of horsepower per weight, miles per dollar, miles per gallon, and size.
The instrument $Z_{ij}$ is defined as the sum of the values of these attributes of other products.

For the purpose of highlighting the effect of clustering assumptions, we report estimates and standard errors under the zero-way cluster robust DML (based on the i.i.d. assumption) and the one-way cluster robust DML (based on clustering along each of the product and market dimensions), as well as the two-way cluster robust DML (along both of the product and market dimensions).
The number $K=4$ of folds of cross fitting is used for the zero- and one-way cluster robust DML, while the number $K^2=4$ of folds of two-way cross fitting is used for the two-way cluster robust DML following the recommendations from Section \ref{sec:simulation_studies} and those by CCDDHNR (\citeyear{CCDDHNR18}, Remark 3.1).
To mitigate the uncertainty induced by sample splitting, we compute estimates based on the average of ten rerandomized DML following CCDDHNR (\citeyear{CCDDHNR18}, Section 3.4) with variance estimation according to CCDDHNR (\citeyear{CCDDHNR18}, Equation 3.13) adapted to our two-way cluster-robustness.

Table \ref{tab:empirical_results} summarizes the results.
For each of the zero-, one-, and two-way cluster robust DML, both the point estimates and standard errors are similar across all the choices of instrument.
Furthermore, the point estimates are also similar across all of the zero-, one-, and two-way cluster robust DML.
On the other hand, the standard errors tend to increase as the assumed number of ways of clustering increases.
In other words, the zero-way cluster robust DML reports the smallest standard error while the two-way cluster robust DML reports the largest standard error.
To robustly account for possible cross-sectional dependence of observations in such two-way cluster sampled data as this market share data, we recommend that researchers use the two-way cluster robust DML although it may incur larger standard errors as is the case with this application.

\section{Conclusion}\label{sec:conclusion}
In this paper, we propose a multiway DML procedure based on a new multiway cross fitting algorithm. This multiway DML procedure is valid in the presence of multiway cluster sampled data, which is frequently used in empirical research.
We present an asymptotic theory showing that multiway DML is valid under nearly identical reguarity conditions to those of CCDDHNR (\citeyear{CCDDHNR18}). 
The proposed method covers a large class of econometric models as is the case with CCDDHNR (\citeyear{CCDDHNR18}), and is compatible with various machine learning based estimation methods. 
Simulation studies indicate that the proposed procedure has attractive finite sample performance under various multiway cluster sampling environments for various machine learning methods. 
To accompany the theoretical findings, we provide easy-to-implement algorithms for multiway DML. 
Such algorithms are readily implementable using existing statistical packages.

There are a couple of possible directions for future research.
First, whereas we focused on linear orthogonal scores that cover a wide range of applications, it may be possible to develop a method and theories for non-linear orthogonal scores as in CCDDHNR (\citeyear{CCDDHNR18}; Section 3.3).
Second, whereas we focused on unconditional moment restrictions, it may be possible and will be important to develop a method and theories for conditional moment restrictions \citep{AiChen2003,AiChen2007,ChenLintonKeilegom2003,ChenPouzo2015}.
We leave these and other extensions for future research.

\newpage
\appendix
\section*{Appendix}
\section{Proofs of the Main Results}

For any $(i,j)\in I_k\times J_\ell$, we use the shorthand notation $\Ep[f(W_{ij})|I_k^c\times J_\ell^c]$ to denote the conditional expectation $\Ep[f(W_{ij})|(W_{i'j'})_{(i',j')\in ([N]\setminus I_k)\times ([M]\setminus J_\ell)}]$ whenever one exists.

\subsection{Proof of Theorem \ref{theorem:main_result_linear}}\label{sec:theorem:main_result_linear}
\begin{proof}
In this proof we try to follow as parallelly as possible the five steps of the proof of Theorem 3.1 of CCDDHNR (\citeyear{CCDDHNR18}) although all the asymptotic arguments are properly modified to account for multiway cluster sampling.  

Denote $\E_n$ for the event $\hat \eta_{k\ell} \in \T_n$ for all $k,\ell\in [K]^2$. Assumption \ref{a:regularity_nuisance_parameters} (i) implies $P(\E_n)\ge 1- K^2\Delta_n$.\\
\noindent \textbf{Step 1.}
This is the main step showing linear representation and asymptotic normality for the proposed estimator. 
Denote 
\begin{align*}
&\hat J:=\frac{1}{K^2}\sumkl \Enkl[\psi^a(W;\hat \eta_{k\ell})],
\qquad R_{n,1}:=\hat J - J_0,\\
&R_{n,2}:=\frac{1}{K^2} \sumkl\Enkl[\psi(W;\theta_0,\hat\eta_{k\ell})]
- \frac{1}{NM} \sumij\psi(W_{ij};\theta_0,\eta_0).
\end{align*}
We will later show in Steps 2, 3, 4 and 5, respectively, that
\begin{align}
&\|R_{n,1}\|=\Opn(\C^{-1/2} + r_n), \label{eq:step2_1}\\
&\|R_{n,2}\|=\Opn(\C^{-1/2}r_n'+ \lambda_n +  \lambda_n'), \label{eq:step2_2}\\
&\Big\|\sqrt{\C}(NM)^{-1} \sumij\psi(W_{ij};\theta_0,\eta_0)\Big\|=\Opn(1), \label{eq:step2_3}\\
&\|\sigma^{-1}\|=\Opn(1). \label{eq:step2_4}
\end{align}
Then,
under Assumptions \ref{a:linear_orthogonal_score} and \ref{a:regularity_nuisance_parameters}, $\C^{-1/2}+r_N\le \rho_n=o(1)$ and all singular values of $J_0$ are bounded away from zero.
Therefore, with $P_n$-probability at least $1-o(1)$, all singular values of  $\hat J$ are bounded away from zero. Thus with the same $P_n$ probability, the multiway DML solution is uniquely written as
\begin{align*}
\tilde \theta =- \hat J^{-1} \frac{1}{K^2}\sumkl \Enkl[\psi^b(W;\hat \eta_{k\ell})],
\end{align*}
and 
\begin{align}
\sqrt{\C}(\tilde \theta - \theta_0)
=&
-\sqrt{\C} \hat J^{-1}\frac{1}{K^2} \sumkl\Big(\Enkl[\psi^b(W;\hat \eta_{k\ell})] +\hat J \theta_0\Big) \nonumber\\
=&-\sqrt{\C} \hat J^{-1}\frac{1}{K^2} \sumkl \Enkl[\psi(W;\theta_0,\hat \eta_{k\ell})] \nonumber\\
=&
-\Big(J_0 + R_{n,1}\Big)^{-1} \times \Big(
\frac{\sqrt{\C}}{NM} \sumij \psi(W_{ij};\theta_0,\eta_0) +\sqrt{\C} R_{n,2}
\Big).
\label{eq:step2_5}
\end{align}
Using the fact that 
\begin{align*}
\Big(J_0 + R_{n,1}\Big)^{-1}  - J_0^{-1}=-(J_0 +R_{n,1})^{-1}R_{n,1} J_0^{-1},
\end{align*}
we have 
\begin{align*}
\|(J_0+R_{n,1})^{-1} - J_0^{-1}\| = &
\|(J_0 +R_{n,1})^{-1}R_{n,1} J_0^{-1}\|\le  \|(J_0 +R_{n,1})^{-1}\|\,\|R_{n,1}\|\, \|J_0^{-1}\|\\
=&\Opn(1)\Opn(\C^{-1/2}+ r_n)\Opn(1)=\Opn(\C^{-1/2}+ r_n).
\end{align*}
Furthermore, $r_n'+\sqrt{\C}(\lambda_n+\lambda_n')\le \rho_n=o(1)$, it holds that 
\begin{align*}
\Big\|\frac{\sqrt{\C}}{NM} \sumij \psi(W_{ij};\theta_0,\eta_0) +\sqrt{\C} R_{n,2}\Big\|
\le&
\Big\|\frac{\sqrt{\C}}{NM} \sumij \psi(W_{ij};\theta_0,\eta_0)\Big\| +\Big\|\sqrt{\C} R_{n,2}\Big\|  \\
=&
\Opn(1) +\opn(1)=\Opn(1),
\end{align*}
where the first equality is due to (\ref{eq:step2_3}) and (\ref{eq:step2_4}).
Combining above two bounds gives 
\begin{align}
\Big\|\Big(J_0 + R_{n,1}\Big)^{-1}  - J_0^{-1}\Big\|\times\Big\|\frac{\sqrt{\C}}{NM} \sumij \psi(W_{ij};\theta_0,\eta_0) +\sqrt{\C} R_{n,2}\Big\| =&
\Opn(\C^{-1/2}+ r_n)\Opn(1) \nonumber\\
=&
\Opn(\C^{-1/2}+ r_n).
\label{eq:step2_6}
\end{align}
Therefore, from (\ref{eq:step2_4}), (\ref{eq:step2_5}) and (\ref{eq:step2_6}), we have
\begin{align*}
\sqrt{\C}\sigma^{-1}(\tilde \theta - \theta_0)=&
\frac{\sqrt{\C}}{NM} \sumij \bar\psi(W_{ij}) +\Opn(\rho_n) .
\end{align*}
The first term on the RHS above can be written as $\GC \bar\psi$. 
Applying Lemma \ref{lemma:hajek}, we obtain the independent linear representation
\begin{align*}
H_n \bar\psi:=\sumi \frac{\sqrt{\C}}{N} \Epn[\bar \psi (W_{ij})|U_{i0}] + \sumj \frac{\sqrt{\C}}{M} \Epn[\bar \psi (W_{ij})|U_{0j}]
\end{align*}
and it holds $P_n$-a.s. that
\begin{align*}
V(\GC \bar\psi)=& V(H_n \bar\psi) + O(\C^{-1})=J_0^{-1}\Gamma (J_0^{-1})' + O(\C^{-1})
\qquad\text{and}\\
\GC \bar\psi =& H_n \bar\psi + \Op(\C^{-1/2})
\end{align*}
under Assumption \ref{a:regularity_nuisance_parameters} (iv).
Recall that $q\ge 4$, the third moments of both summands of $H_n \bar\psi$ are bounded over $n$ under Assumptions \ref{a:linear_orthogonal_score}(v) and \ref{a:regularity_nuisance_parameters} (ii) (iv).
We have verified all the conditions for Lyapunov's CLT. An application of Lyapunov's CLT and Cramer-Wold device gives
\begin{align*}
&H_n \bar\psi \leadsto N(0,I_{d_\theta})
\end{align*}
and an application of Theorem 2.7 of \cite{vdV98} concludes the proof.

\noindent \textbf{Step 2.}
Since $K$ is fixed, it suffices to show for any $(k,\ell)\in [K]^2$,
\begin{align*}
\Big\|\Enkl[\psi^a(W;\hat \eta_{k\ell})]-\Ep[\psi^a(W_{11};\eta_0)]\Big\|
=
\Opn(\C^{-1/2} + r_n).
\end{align*}
Fix $(k,\ell)\in [K]^2$,
\begin{align*}
&\Big\|\Enkl[\psi^a(W;\hat \eta_{k\ell})]-\Epn[\psi^a(W_{ij};\eta_0)]\Big\|
\le \mathcal I_{1,k\ell} + \mathcal I_{2,k\ell}.
\end{align*}
where
\begin{align*}
\mathcal I_{1,k\ell} &:= \Big\|\Enkl[\psi^a(W;\hat \eta_{k\ell})]-\Epn[\psi^a(W_{ij};\hat \eta_{k\ell})|I_k^c \times J_\ell^c]\Big\|
\\
\mathcal I_{2,k\ell} &:= \Big\|\Epn[\psi^a(W_{ij};\hat \eta_{k\ell})|I_k^c \times J_\ell^c]-\Epn[\psi^a(W_{11};\eta_0)]\Big\|.
\end{align*}
Notice that $\mathcal I_{2,k\ell}\le r_n$ with $P_n$-probability $1-o(1)$ follows directly from Assumptions \ref{a:sampling} (ii) and \ref{a:regularity_nuisance_parameters} (iii). 
Now denote $\tilde\psi^a_{ij,m}=\psi^a_m(W_{ij};\hat \eta_{k\ell})-\Epn[\psi^a_m(W_{ij};\hat \eta_{k\ell})|I_k^c\times J_\ell^c]$ and $\tilde \psi^a_{ij}=(\tilde \psi^a_{ij,m})_{m\in [d_\theta]}$. To bound $\mathcal I_{1,k\ell}$, note that conditional on $I_k^c\times J_\ell^c$, it holds that
\begin{align*}
\Epn[\mathcal I_{1,k\ell}^2|I_k^c\times J_\ell^c]
=&
\Epn\Big[\Big\|\Enkl[\psi^a(W;\hat \eta_{k\ell})]-\Epn[\psi^a(W_{ij};\hat \eta_{k\ell})|I_k^c\times J_\ell^c]\Big\|^2\Big|I_k^c\times J_\ell^c\Big]\\
=&
\frac{1}{(|I||J|)^2}\Epn\Big[\sum_{m=1}^{d_\theta}\Big(\sumIJ\tilde \psi^a_{ij,m}\Big)^2\Big|I_k^c\times J_\ell^c\Big]\\
=&
\frac{1}{(|I||J|)^2}\sumIJ\sum_{j'\in J_\ell,j'\ne j}\Epn\Big[\sum_{m=1}^{d_\theta}\tilde \psi^a_{ij,m}\tilde \psi^a_{ij',m}\Big|I_k^c\times J_\ell^c\Big]\\
&+
\frac{1}{(|I||J|)^2}\sumIJ\sum_{i'\in I_k,i'\ne i}\Epn\Big[\sum_{m=1}^{d_\theta}\tilde \psi^a_{ij,m}\tilde \psi^a_{i'j,m}\Big|I_k^c\times J_\ell^c\Big]
\\
&+
\frac{1}{(|I||J|)^2}\sumIJ\Epn\Big[\sum_{m=1}^{d_\theta}(\tilde \psi^a_{ij,m})^2\Big|I_k^c\times J_\ell^c\Big]
+ 0
\\
= &
\frac{1}{(|I||J|)^2} \sumIJ\sum_{j'\in J_\ell,j'\ne j}\Epn[\langle\tilde\psi^a_{ij},\tilde\psi^a_{ij'}\rangle|I_k^c\times J_\ell^c]\\
&+
\frac{1}{(|I||J|)^2} \sumIJ\sum_{i'\in I_k,i'\ne i}\Epn[\langle\tilde\psi^a_{ij},\tilde\psi^a_{i'j}\rangle|I_k^c\times J_\ell^c]\\
&+\frac{1}{(|I||J|)^2} \sumIJ\Epn[\|\tilde\psi^a_{ij}\|^2|I_k^c\times J_\ell^c]
\\
\lesssim&
\frac{1}{|I|\wedge|J|} \Epn\Big[\Big\|\psi^a(W_{ij};\hat\theta_{k\ell})-\Epn[\psi^a(W_{ij};\hat\theta_{k\ell})|I_k^c\times J_\ell^c]\Big\|^2\Big|I_k^c\times J_\ell^c\Big] \\
\le&
\frac{1}{|I|\wedge|J|} \Epn[\|\psi^a(W_{ij};\hat\theta_{k\ell})\|^2|I_k^c\times J_\ell^c] \\
\le& c_1^2/|I|\wedge |J|
\end{align*}
\noindent under an application of Cauchy-Schwartz's inequality and Assumptions \ref{a:sampling} and \ref{a:regularity_nuisance_parameters} (ii). Note that $\C\lesssim |I|\wedge|J|\lesssim\C$. Hence an application of Lemma \ref{lemma:conditional_convergence} (i) implies 
$
\mathcal I_{1,k\ell}=\Opn(\C^{-1/2}).
$
This completes a proof of (\ref{eq:step2_1}).

\noindent \textbf{Step 3.}
It again suffices to show that for any $(k,\ell)\in [K]^2$, one has
\begin{align*}
\Big\|\Enkl[\psi(W;\theta_0,\hat\eta_{k\ell})]
- \frac{1}{|I| |J|}\sumIJ\psi(W_{ij};\theta_0,\eta_0)\Big\|=\Opn(\C^{-1/2}r_n'+ \lambda_n +  \lambda_n')
\end{align*}
Denote 
$$\Gnkl[\phi(W)]=\frac{\sqrt{\C}}{|I||J|}\sumIJ \Big( \phi(W_{ij}) - \int \phi(w) dP_n \Big), $$
where $\phi$ is $P_n$ an integrable function on $\supp(W)$.
Then 
\begin{align*}
&\Big\|\Enkl[\psi(W;\theta_0,\hat\eta_{k\ell})]
- \frac{1}{|I| |J|} \sumIJ\psi(W_{ij};\theta_0,\eta_0)\Big\|
\le
\frac{\mathcal I_{3,k\ell}+\mathcal I_{4,k\ell}}{\sqrt{\C}}
\end{align*}
where
\begin{align*}
\mathcal I_{3,k\ell}
:=&
\big\|\Gnkl[\psi(W;\theta_0,\hat \eta_{k,\ell})]-\Gnkl[\psi(W;\theta_0,\eta_0)]\big\|,
\\
\mathcal I_{4,k\ell}
:=&\sqrt{\C}
\Big\|
\Epn[\psi(W_{ij};\theta_0,\hat \eta_{k,\ell})|I_k\times J_\ell]- \Epn[\psi(W_{11};\theta_0,\eta_0)]
\Big\|.
\end{align*}
Denote $\tilde \psi_{ij,m}:=\psi_m(W_{ij};\theta_0,\hat \eta_{k,\ell})-\psi_m(W_{ij};\theta_0,\eta_0)$ and $\tilde \psi_{ij}=(\tilde \psi_{ij,m})_{m\in[d_\theta]}$. To bound $\mathcal I_{3,k\ell}$, notice that using a similar argument as for the bound of $\mathcal I_{1,k\ell}$, one has
{\small
\begin{align*}
\Epn[\|\mathcal I_{3,k\ell}\|^2|I_k^c\times J_\ell^c]
=&\Epn[\|\Gnkl[\psi(W_{ij};\theta_0,\hat \eta_{k,\ell})-\psi(W_{ij};\theta_0,\eta_0)]\|^2|I_k^c\times J_\ell^c]\\
=&\Epn\Big[\frac{\C}{(|I||J|)^2}\sum_{m=1}^{d_\theta}\Big\{\sumIJ \Big(\tilde\psi_{ij,m}
-\Epn\tilde\psi_{ij,m}
\Big)\Big\}^2\Big|I_k^c\times J_\ell^c\Big]\\
=&
\frac{\C}{(|I||J|)^2}\sumIJ\sum_{j'\in J_\ell, j'\ne j} \Epn\Big[\sum_{m=1}^{d_\theta}\Big(\tilde\psi_{ij,m}-\Epn\tilde\psi_{ij,m}
\Big)\Big(\tilde\psi_{ij',m}-\Epn\tilde\psi_{ij',m}
\Big)\Big|I_k^c\times J_\ell^c\Big]\\
&+
\frac{\C}{(|I||J|)^2}\sumIJ\sum_{i'\in I_k, i'\ne i} \Epn\Big[\sum_{m=1}^{d_\theta}\Big(\tilde\psi_{ij,m}-\Epn\tilde\psi_{ij,m}
\Big)\Big(\tilde\psi_{i'j,m}-\Epn\tilde\psi_{i'j,m}
\Big)\Big|I_k^c\times J_\ell^c\Big]\\
&+
\frac{\C}{(|I||J|)^2}\sumIJ \Epn\Big[\sum_{m=1}^{d_\theta}\Big(\tilde\psi_{ij,m}-\Epn\tilde\psi_{ij,m}
\Big)^2\Big|I_k^c\times J_\ell^c\Big]
+0\\
=&
\frac{\C}{(|I||J|)^2}\sumIJ\sum_{j'\in J_\ell, j'\ne j} \Epn\Big[\langle\tilde\psi_{ij}-\Epn\tilde\psi_{ij}
,\tilde\psi_{ij'}-\Epn\tilde\psi_{ij'}
\rangle\Big|I_k^c\times J_\ell^c\Big]\\
&+
\frac{\C}{(|I||J|)^2}\sumIJ\sum_{i'\in I_k, i'\ne i} \Epn\Big[\langle\tilde\psi_{ij}-\Epn\tilde\psi_{ij}
,\tilde\psi_{i'j}-\Epn\tilde\psi_{i'j}
\rangle\Big|I_k^c\times J_\ell^c\Big]\\
&+
\frac{\C}{(|I||J|)^2}\sumIJ \Epn\Big[\Big\|\tilde\psi_{ij}-\Epn\tilde\psi_{ij}
\Big\|^2\Big|I_k^c\times J_\ell^c\Big]\\
\lesssim&
 \Epn\Big[\Big\|\psi(W_{ij};\theta_0,\hat\eta)-\psi(W_{ij};\theta_0,\eta_0)-\Epn[\psi(W_{ij};\theta_0,\hat\eta)-\psi(W_{ij};\theta_0,\eta_0)]\Big\|^2\Big|I_k^c\times J_\ell^c\Big] \\
\leq&
 \Epn[\|\psi(W_{ij};\theta_0,\hat\eta)-\psi(W_{ij};\theta_0,\eta_0)\|^2|I_k^c\times J_\ell^c] \\
 \le&
 \sup_{\eta \in \T_n}\Epn[\|\psi(W_{00};\theta_0,\eta)-\psi(W_{00};\theta_0,\eta_0)\|^2|I_k^c\times J_\ell^c] \\
 =&
  \sup_{\eta \in \T_n}\Epn[\|\psi(W_{00};\theta_0,\eta)-\psi(W_{00};\theta_0,\eta_0)\|^2] = (r_n')^2,
\end{align*}}
where the first inequality follows from Cauchy-Schwartz's inequality, the second-to-last equality is due to Assumption \ref{a:sampling}, and the last equality is due to Assumption \ref{a:regularity_nuisance_parameters} (iii).

Hence, $\mathcal I_{3,k\ell}=\Opn(r_n')$.
To bound $\mathcal I_{4,k\ell}$, let
\begin{align*}
f_{k\ell}(r):=
\Epn[\psi(W_{ij};\theta_0,\eta_0 + r(\hat \eta_{k\ell}-\eta_0))|I_k^c\times J_\ell^c]
-
\Epn[\psi(W_{11};\theta_0,\eta_0)],
\qquad r\in[0,1].
\end{align*}
An application of the mean value expansion coordinate-wise gives
\begin{align*}
f_{k\ell}(1)=f_{k\ell}(0)+f_{k\ell}'(0)+f_{k\ell}''(\tilde r)/2,
\end{align*}
where $\tilde r \in (0,1)$.
Note that $f_{k\ell}(0)=0$ under Assumption \ref{a:linear_orthogonal_score} (i), and
\begin{align*}
\|f_{k\ell}'(0)\|=\Big\| \partial_\eta \Epn \psi(W;\theta_0,\eta_0)[\hat \eta_{k\ell}-\eta_0] \Big\|\le \lambda_n
\end{align*}
under Assumption \ref{a:linear_orthogonal_score} (iv).
Moreover, under Assumption \ref{a:regularity_nuisance_parameters} (iii), on the event $\E_n$, we have
\begin{align*}
\|f_{k\ell}''(\tilde r)\|\le \sup_{r\in(0,1)}\|f_{k\ell}''(r)\|\le \lambda_n'.
\end{align*}
This completes a proof of (\ref{eq:step2_2}).

\noindent \textbf{Step 4.}
Note that
\begin{align*}
\Epn\Big[\Big\|\frac{\sqrt{\C}}{NM}\sumij\psi(W_{ij};\theta_0,\eta_0)\Big\|^2\Big]
=&
\frac{\C}{(NM)^2}\Epn\Big[\sum_{m=1}^{d_\theta}\Big(\sumij\psi_m(W_{ij};\theta_0,\eta_0)\Big)^2\Big]\\
=&
\frac{\C}{(NM)^2}\sum_{i=1}^N\sum_{1\le j< j'\le M}\Epn\Big[\sum_{m=1}^{d_\theta}\psi_m(W_{ij};\theta_0,\eta_0)\psi_m(W_{ij'};\theta_0,\eta_0)\Big]\\
&+
\frac{\C}{(NM)^2}\sum_{1\le i<i'\le N}\sum_{j=1}^M\Epn\Big[\sum_{m=1}^{d_\theta}\psi_m(W_{ij};\theta_0,\eta_0)\psi_m(W_{i'j};\theta_0,\eta_0)\Big]\\
&+
\frac{\C}{(NM)^2}\sumij\Epn\Big[\sum_{m=1}^{d_\theta}\psi^2_m(W_{ij};\theta_0,\eta_0)\Big] + 0\\
\lesssim& 
\Epn[\|\psi(W_{ij};\theta_0,\eta_0)\|^2]\le c_1^2
\end{align*}
under Assumptions \ref{a:sampling} and \ref{a:regularity_nuisance_parameters} (ii).
Therefore, an application of Markov's inequality implies
\begin{align*}
\Big\|\frac{\sqrt{\C}}{NM} \sumij\psi(W_{ij};\theta_0,\eta_0)\Big\|=\Opn(1).
\end{align*}
This completes a proof of (\ref{eq:step2_3}).

\noindent \textbf{Step 5.}
Note that all singular values of $J_0$ are bounded from above by $c_1$ under Assumption \ref{a:linear_orthogonal_score} (v) and all eigenvalues of $\Gamma$ are bounded from below by $c_0$ under Assumption \ref{a:regularity_nuisance_parameters} (iv). 
Therefore, we have $\|\sigma^{-1}\|\le c_1/\sqrt{c_0}$ and thus
$
\|\sigma^{-1}\|
=\Opn(1).
$
This completes a proof of (\ref{eq:step2_4}).
\end{proof}

\subsection{Proof of Theorem \ref{theorem:variance_estimator_linear}}\label{sec:theorem:variance_estimator_linear}
\begin{proof}
Step 2 of the proof of Theorem \ref{theorem:main_result_linear} proves $\|\hat J - J_0\| = O_p(\C^{-1/2} + r_n)$ and Assumption \ref{a:linear_orthogonal_score} (v) implies $\|J_0^{-1}\| \leq c_0^{-1}$.
Therefore, to prove the claim of the theorem, it suffices to show
\begin{align*}
& \ \Big\|\frac{1}{K^2}\sumkl\Big\{
\frac{|I|\wedge|J|}{(|I||J|)^2}\sum_{i\in I_k}\sum_{j,j'\in J_\ell} \psi(W_{ij};\tilde \theta,\hat\eta_{k\ell}) \psi(W_{ij'};\tilde \theta,\hat\eta_{k\ell})'
\\
&\qquad\qquad\qquad +
\frac{|I|\wedge|J|}{(|I||J|)^2}\sum_{i,i'\in I_k}\sum_{j\in J_\ell} \psi(W_{ij};\tilde \theta,\hat\eta_{k\ell}) \psi(W_{i'j};\tilde \theta,\hat\eta_{k\ell})' 
\Big\}
\\
&-
\bar\mu_N\Ep [\psi(W_{11};\theta_0,\eta_0)\psi(W_{12};\theta_0,\eta_0)'] 
- \bar\mu_M\Ep[\psi(W_{11};\theta_0,\eta_0)\psi(W_{21};\theta_0,\eta_0)']
\Big\|=\Op(\rho_n).
\end{align*}
Moreover, since $K$ and $d_{\theta}$ are constants and $\mu_N\to \bar \mu_N\le 1$ and $\mu_M\to \bar \mu_M\le 1$, it suffices to show that for each $(k,\ell)\in [K]^2$ and $l,m\in[d_\theta]$, it holds that 
\begin{align*}
&\Big|
\frac{|I|\wedge|J|}{(|I||J|)^2}\sum_{i\in I_k}\sum_{j,j'\in J_\ell} \psi_l(W_{ij};\tilde \theta,\hat\eta_{k\ell}) \psi_m(W_{ij'};\tilde \theta,\hat\eta_{k\ell})-\mu_N\Ep [\psi_l(W_{11};\theta_0,\eta_0)\psi_m(W_{12};\theta_0,\eta_0)] \Big|=\Op(\rho_n)
\end{align*}
and
\begin{align*}
&\Big|\frac{|I|\wedge|J|}{(|I||J|)^2}\sum_{i,i'\in I_k}\sum_{j\in J_\ell} \psi_l(W_{ij};\tilde \theta,\hat\eta_{k\ell}) \psi_m(W_{i'j};\tilde \theta,\hat\eta_{k\ell})
- 
\mu_M\Ep[\psi_l(W_{11};\theta_0,\eta_0)\psi_m(W_{21};\theta_0,\eta_0)]
\Big|=\Op(\rho_n).
\end{align*}

We will show the second statement since the first one follows analogously. 
Denote the left-hand side of the equation as $\mathcal I_{k\ell,lm}$. 
First, note that $(|I|\wedge|J|)/|J|=\mu_M$, and apply the triangle inequality to get
\begin{align*}
\mathcal I_{k\ell,lm}\le \mathcal I_{k\ell,lm,1}+ \mathcal I_{k\ell,lm,2},
\end{align*}
where
\begin{align*}
&\mathcal I_{k\ell,lm,1}:=
\Big|\frac{1}{|I|^2|J|}\sum_{i,i'\in I_k}\sum_{j\in J_\ell} \Big\{\psi_l(W_{ij};\tilde \theta,\hat\eta_{k\ell}) \psi_m(W_{i'j};\tilde \theta,\hat\eta_{k\ell})
-
 \psi_l(W_{ij};\theta_0,\eta_0) \psi_m(W_{i'j};\theta_0,\eta_0) \Big\}
\Big|\\
&\mathcal I_{k\ell,lm,2}:=
\Big|\frac{1}{|I|^2|J|}\sum_{i,i'\in I_k}\sum_{j\in J_\ell} \psi_l(W_{ij};\theta_0,\eta_0) \psi_m(W_{i'j};\theta_0,\eta_0)
-
\Ep [\psi_l(W_{11};\theta_0,\eta_0)\psi_m(W_{21};\theta_0,\eta_0)] 
\Big|.
\end{align*}
We first find a bound for $\mathcal I_{k\ell,lm,2}$. Since $q>4$, it holds that
\begin{align*}
\Ep[\mathcal I_{k\ell,lm,2}^2]
=&
\frac{1}{|I|^4|J|^2}\Ep\Big[
\Big|\sum_{i,i'\in I_k}\sum_{j\in J_\ell} \psi_l(W_{ij};\theta_0,\eta_0) \psi_m(W_{i'j};\theta_0,\eta_0)
-
\Ep [\psi_l(W_{11};\theta_0,\eta_0)\psi_m(W_{21};\theta_0,\eta_0)] 
\Big|^2
\Big]\\
\le &
\frac{1}{|I|^4|J|^2}\Ep\Big[
\sum_{i,i',i''\in I_k}\sum_{j,j'\in J_\ell} \psi_l(W_{ij};\theta_0,\eta_0) \psi_m(W_{i'j};\theta_0,\eta_0)
\psi_l(W_{ij'};\theta_0,\eta_0) \psi_m(W_{i''j'};\theta_0,\eta_0)
\Big]\\
&+\frac{1}{|I|^4|J|^2}\Ep\Big[
\sum_{i,i',i'',i'''\in I_k}\sum_{j\in J_\ell} \psi_l(W_{ij};\theta_0,\eta_0) \psi_m(W_{i'j};\theta_0,\eta_0)\psi_l(W_{i''j};\theta_0,\eta_0) \psi_m(W_{i'''j};\theta_0,\eta_0)
\Big]\\
&+o((|I|\wedge |J|)^{-1}) + 0\\
\lesssim&\frac{1}{|I|\wedge |J|}\Ep[\|\psi(W;\theta_0,\eta_0)\|^4]\lesssim c_1^4/\C=O(\C^{-1/2}).
\end{align*}

Now, to bound $\mathcal I_{k\ell,lm,1}$, we make use of the following identity coming from the proof of Theorem 3.2 in CCDDHNR (\citeyear{CCDDHNR18}): for any numbers $a$, $b$, $\delta a$, $\delta b$ such that
$|a|\vee |b|\le c$ and $|\delta a |\vee |\delta b| \le r$, it holds that 
$
|(a+\delta a)(b+\delta b)-ab|\le 2r( c+r).
$ Denote $\psi_{ij,h}:=\psi_l(W_{ij};\theta_0,\eta_0)$ and $\hat\psi_{ij,h}:=\psi_l(W_{ij};\tilde \theta,\hat\eta_{k\ell})$ for $h\in\{l,m\}$ and apply the above identity with $a=\psi_{ij,l}$, $b=\psi_{i'j,m}$, $a+\delta a =\hat\psi_{ij,l}$, $b+ \delta b=\hat\psi_{i'j,m}$, $r=|\hat\psi_{ij,l}-\psi_{ij,l}|\vee |\hat\psi_{i'j,m}-\psi_{i'j,m}|$ and $c=|\psi_{ij,l}|\vee|\psi_{i'j,m}|$.
Then
\begin{align*}
\mathcal I_{k\ell,lm,1}=&
\Big|\frac{1}{|I|^2|J|}\sum_{i,i'\in I_k}\sum_{j\in J_\ell} \Big\{
\hat\psi_{ij,l} \hat\psi_{i'j,m}
-
\psi_{ij,l} \psi_{i'j,m}  \Big\}
\Big|\\
\le &
\frac{1}{|I|^2|J|}\sum_{i,i'\in I_k}\sum_{j\in J_\ell}|
\hat\psi_{ij,l} \hat\psi_{i'j,m}
-
\psi_{ij,l} \psi_{i'j,m} |\\
\le& 
\frac{2}{|I|^2|J|}\sum_{i,i'\in I_k}\sum_{j\in J_\ell} (|\hat\psi_{ij,l}-\psi_{ij,l}|\vee |\hat\psi_{i'j,m}-\psi_{i'j,m}|)
\\
&\qquad \times \Big(|\psi_{ij,l}|\vee|\psi_{i'j,m}|
+|\hat\psi_{ij,l}-\psi_{ij,l}|\vee |\hat\psi_{i'j,m}-\psi_{i'j,m}|\Big)\\
\le& 
\Big(\frac{2}{|I|^2|J|}\sum_{i,i'\in I_k}\sum_{j\in J_\ell} |\hat\psi_{ij,l}-\psi_{ij,l}|^2\vee |\hat\psi_{i'j,m}-\psi_{i'j,m}|^2\Big)^{1/2}\\
&\qquad\times
 \Big(\frac{2}{|I|^2|J|}\sum_{i,i'\in I_k}\sum_{j\in J_\ell}\Big\{|\psi_{ij,l}|\vee|\psi_{i'j,m}|
+|\hat\psi_{ij,l}-\psi_{ij,l}|\vee |\hat\psi_{i'j,m}-\psi_{i'j,m}|\Big\}^2\Big)^{1/2}\\
\le&
\Big(\frac{2}{|I|^2|J|}\sum_{i,i'\in I_k}\sum_{j\in J_\ell} |\hat\psi_{ij,l}-\psi_{ij,l}|^2\vee |\hat\psi_{i'j,m}-\psi_{i'j,m}|^2\Big)^{1/2}\\
&\times\Big\{
 \Big(\frac{2}{|I|^2|J|}\sum_{i,i'\in I_k}\sum_{j\in J_\ell}|\psi_{ij,l}|^2\vee|\psi_{i'j,m}|^2\Big)^{1/2} 
\\
&
+
  \Big(\frac{2}{|I|^2|J|}\sum_{i,i'\in I_k}\sum_{j\in J_\ell} |\hat\psi_{ij,l}-\psi_{ij,l}|^2\vee |\hat\psi_{i'j,m}-\psi_{i'j,m}|^2\Big)^{1/2}\Big\},
\end{align*}
where the second to the last inequality follows the Cauchy-Schwartz's inequality and Minkowski's inequality.
Notice that
\begin{align*}
&\sum_{i,i'\in I_k}\sum_{j\in J_\ell}|\psi_{ij,l}|^2\vee|\psi_{i'j,m}|^2\le   |I|\sumij \|\psi(W_{ij};\theta_0,\eta_0)\|^2,\\
&\sum_{i,i'\in I_k}\sum_{j\in J_\ell}|\hat \psi_{ij,l}-\psi_{ij,l}|^2\vee|\hat \psi_{i'j,m}-\psi_{i'j,m}|^2\le |I|\sumij \|\psi(W_{ij};\tilde\theta,\hat \eta_{k\ell})-\psi(W_{ij};\theta_0,\eta_0)\|^2.
\end{align*}
Thus, the above bound for $\mathcal I_{k\ell,lm,1}$ implies that
\begin{align*}
\mathcal I_{k\ell,lm,1}^2
\lesssim&
R_n\times \Big(\frac{1}{|I||J|}\sumIJ
\| \psi(W_{ij};\theta_0, \eta_0) 
\|^2
+
R_n
\Big),
\end{align*}
where
\begin{align*}
R_n:=\frac{1}{|I||J|}\sumIJ
\| \psi(W_{ij};\tilde\theta,\hat \eta_{k\ell}) - \psi(W_{ij};\theta_0,\eta_0) 
\|^2.
\end{align*}
Notice that 
\begin{align*}
\frac{1}{|I||J|}\sumIJ
\| \psi(W_{ij};\theta_0,\eta_0) 
\|^2=\Op(1),
\end{align*}
which is implied by Markov's inequality and the calculations
\begin{align*}
\Ep\Big[\frac{1}{|I||J|}\sumIJ
\| \psi(W_{ij};\theta_0,\eta_0) 
\|^2\Big]=&\Ep[ \|\psi(W_{11};\theta_0,\eta_0) 
\|^2]\le c_1^2
\end{align*}
under Assumptions \ref{a:sampling} and \ref{a:regularity_nuisance_parameters} (ii). 
Finally, to bound $R_n$, using Assumption \ref{a:linear_orthogonal_score} (ii), 
\begin{align*}
R_n\lesssim&
 \frac{1}{|I||J|}\sumIJ
\| \psi^a(W_{ij};\hat \eta_{k\ell})(\tilde \theta -\theta_0)
\|^2 +
\frac{1}{|I||J|}\sumIJ
\| \psi(W_{ij};\theta_0,\hat\eta_{k\ell}) -\psi(W_{ij};\theta_0,\eta_0) 
\|^2.
\end{align*}
The first term on RHS is bounded by
\begin{align*}
 \Big(\frac{1}{|I||J|}\sumIJ
\| \psi^a(W_{ij};\hat \eta_{k\ell})
\|^2 \Big)\times\|\tilde \theta -\theta_0\|^2=\Op(1)\times \Op(\C^{-1})=\Op(\C^{-1})
\end{align*}
due to Assumption \ref{a:regularity_nuisance_parameters} (ii), Markov's inequality, and Theorem \ref{theorem:main_result_linear}.
Furthermore, given that $(W_{ij})_{(i,j)\in I_k^c \times J_\ell^c}$ satisfies $\hat \eta_{k\ell}\in\mathcal T_n$,
\begin{align*}
\Ep\Big[ \|\psi(W_{ij};\theta_0,\hat\eta_{k\ell}) -\psi(W_{ij};\theta_0,\eta_0)\|^2\Big|I_k^c \times J_\ell^c\Big]\le&
\sup_{\eta\in \mathcal T_n}\Ep\Big[ \|\psi(W_{ij};\theta_0,\eta) -\psi(W_{ij};\theta_0,\eta_0)\|^2\Big|I_k^c \times J_\ell^c\Big] \le (r_n')^2
\end{align*}
due to Assumptions \ref{a:sampling} and \ref{a:regularity_nuisance_parameters} (iii).
Also, the event $\hat \eta_{k\ell}\in\mathcal T_n$ happens with probability $1-o(1)$, we have $R_n=\Op(\C^{-1}+(r'_n)^2)$. Thus we conclude that 
\begin{align*}
\mathcal I_{k\ell,lm,1}=\Op(\C^{-1/2}+r'_n).
\end{align*}
This completes the proof.
\end{proof}

\section{Useful Lemmas}
We collect some of the useful auxiliary results in this section.

First, for any $f:\supp(W)\to  \Real^d$ for a fixed $d\in \mathbbm N$, we use
\begin{align*}
\GC f:=\sqrt{\C}\Big\{\frac{1}{NM} \sumi\sumj f(W_{ij}) - \Ep[f(W_{11})]\Big\}
\end{align*}
to denote its multiway empirical process. 
The following is a multivariate version of \cite{ChiangSasaki2019}, Lemma 1; see also Lemma D.2 in \cite{DDG18}.
\begin{lemma}[Independentization via H\'ajek Projections]\label{lemma:hajek}
If Assumption \ref{a:sampling} holds and $f:\supp(W)\to  \Real^d$ for some fixed $d\in \mathbbm N$ and suppose $\Ep \|f(W_{11})\|^2<K$ for a finite constant $K$ that is independent of $n$, then there exist i.i.d. uniform random variables $U_{i0}$ and $U_{0j}$ such that the H\'ajek projection $H_n f$ of $\GC f$ on 
$$
\G_n=\Big\{ \sumi g_{i0}(U_{i0}) + \sumj g_{0j}(U_{0j}) : g_{i0}, g_{0j} \in L^2(P_{n}) \Big\}
$$
is equal to
\begin{align*}
H_n f=\frac{\sqrt{\C}}{N}\sumi   \Ep\Big[f(W_{i1})- \Ep f(W_{11}) \Big| U_{i0}\Big] +
\frac{\sqrt{\C}}{M} \sumj \Ep\Big[f(W_{1j})- \Ep f(W_{11}) \Big| U_{0j}\Big]
\end{align*}
for each $n$.
Furthermore,
\begin{align*}
V(\GC f)= V(H_n f)+O(\C^{-1})=\bar\mu_N  Cov(f(W_{11}),f(W_{12})) + \bar\mu_M Cov(f(W_{11}),f(W_{21}))+O(\C^{-1})
\end{align*}
holds a.s.
\begin{proof}
The proof is essentially the same as the proof for Lemma 1 of \cite{ChiangSasaki2019} and is therefore omitted.
\end{proof}
\end{lemma}
The following re-states Lemma 6.1. of CCDDHNR (\citeyear{CCDDHNR18}):
\begin{lemma}[Conditional Convergence Implies Unconditional]\label{lemma:conditional_convergence}
Let $(X_n)$ and $(Y_n)$ be sequences of random vectors.
\begin{enumerate}[(i)]
\item If for $\epsilon_n\to 0$, $P(\|X_n\|>\epsilon_n|Y_n)=\op(1)$ in probability, then $P(\|X_n\|>\epsilon_n)=o(1)$. In particular, this occurs if $\Ep[\|X_n\|^q/\epsilon_n^q|Y_n]=\op(1)$ for some $q\ge 1$.
\item Let $(A_n)$ be a sequence of positive constants. If $\|X_n\|=\Op(A_n)$ conditional on $Y_n$, then $\|X_n\|=\Op(A_n)$ unconditional, namely, for any $l_n\to \infty$, 
$P(\|X_n\|>l_n A_n)=o(1)$.
\end{enumerate}
\end{lemma}

\section{Extension to General Multiway Clustering}\label{sec:extension_to_general_multiway_clustering}
In this section, we extend the main results to general multiway cluster sampling framework.
Notations in the current section are independent of those in the remaining parts of the paper -- we introduce different notations in order to enhance the readability of the main results of the paper while economizing complicated notations in the current extension section.
Consider the $\ell$-way clustered data for a fixed dimension $\ell \in \mathbb{N}$.
With $C_i \in \mathbb{N}$ denoting the number of clusters in the $i$-th cluster dimension for each $i \in \{1,...,\ell\}$,
each cell of the $\ell$-way clustered sample is indexed by the $\ell$-dimensional multiway cluster indices $\emph{\textbf{j}}=(j_1,...,j_\ell) \in \times_{i=1}^\ell [C_i]$.
The $\ell$-dimensional size $(C_1,...,C_\ell)\in \mathbb{N}^\ell$ of the $\ell$-way clustered sample will be index by $n\in\mathbb{N}$ as $(C_1,...,C_\ell)=(C_1(n),...,C_\ell(n))$, where $C_i(n)$ is non-decreasing in $n$ for each $i \in \{1,...,\ell\}$ and $\prod\limits_{i=1}^\ell C_i(n)$ is increasing in $n$. 
With this said, we will suppress the index notation and write $(C_1,...,C_\ell)$ without $n$ for simplicity.
Also define the notations $\emph{\textbf{C}}=(C_1,...,C_\ell)$, $\prod_C=\prod\limits_{i=1}^\ell C_i$, $\underline{C}=\min_{1\leq i \leq \ell}C_i$, $\overline{C}=\max_{1\leq i \leq \ell}C_i$, and $\mu_i=\underline{C}/C_i$ for each $i \in \{1,...,\ell\}$. 
Suppose that $\mu_i \to \bar{\mu_i}$ for some constant $\bar{\mu_i}$ for each $i \in \{1,...,\ell\}$.
The number of observations in the $\emph{\textbf{j}}$-th cell is denoted by $N_\emph{\textbf{j}}$, which is treated as an $\{0,1,...,\overline{N}\}$-valued random variable for some $\overline{N} \in \mathbb{N}$ not depending on $n$. 
When $[\cdot]$ takes the random variable $N_\emph{\textbf{j}}$ as an argument, we extend the definition of $[\cdot]$ to $[N_\emph{\textbf{j}}] := \{1,...,N_\emph{\textbf{j}}\}$ if $N_\emph{\textbf{j}} \ge 1$ and $:= \emptyset$ if $N_\emph{\textbf{j}} = 0$. 
The observed vector for unit $\imath \in [N_\emph{\textbf{j}}]$ in the $\emph{\textbf{j}}$-th cell is denoted by $W_{\imath, \emph{\textbf{j}}}$.
Let $\{\mathcal{P}_n\}_n$ be a sequence of sets of probability laws of $(N_\emph{\textbf{j}},(W_{\imath,\emph{\textbf{j}}})_{1\leq \imath \leq \overline{N}})_{\emph{\textbf{j}}\geq \textbf{1}}$, where $\mathbf{1}:=(1,...,1)$ for a short-hand notation and we write $\emph{\textbf{j}}\geq\emph{\textbf{j}}^{\:\prime}$ to mean $j_i\geq j_i{'}$ for all $i \in \{1,...,\ell\}$. 

\begin{example}\label{ex:overview}
The sampling setting in Section \ref{sec:setup} fits in the current general framework with $\ell = 2$, $C_1 = N$, $C_2 = M$, and $(N_{\emph{\textbf{j}}},W_{1,\emph{\textbf{j}}}) = (1,W_{j_1 j_2})$ for all $\emph{\textbf{j}} \in [N] \times [M]$ with probability one. $\triangle$
\end{example}

The econometric model has the true parameters $(\theta_0,\eta_0) \in \Theta \times T$ satisfying the score moment restriction 
\begin{align}
\Ep\Big[\sum\limits_{\imath=1}^{N_\textbf{1}}\psi( W_{\imath,\textbf{1}};\theta_0,\eta_0)\Big]=0,
\label{eq:existance_condition 2}
\end{align} 
where we focus on the linear Neyman orthogonal score of the form
\begin{align}
\psi(w;\theta,\eta)=\psi^a(w;\eta)\theta +\psi^b(w;\eta), \text{ for all $w\in \supp(W)$, $\theta\in\Theta$, $\eta\in T$ } \label{eq:linear_score2}
\end{align}
for supp(W)$:=\cup_{\imath=1}^{\overline N}$supp$(W_{\imath,\textbf{1}})$, $\Theta \subset \mathbb{R}^{d_\theta}$ and a convex set $T$.

For a fixed integer $K>1$, we randomly split the data into $K$ folds in each of the $\ell$ cluster dimensions, resulting in $K^\ell$ folds in total.
Specifically, randomly partition $[C_i]$ into $K$ parts $\{I_i^1,...,I_i^K\}$ for each $i \in \{1,...,\ell\}$.
We use the $\ell$-dimensional indices $\emph{\textbf{k}}:=(k_1,...,k_\ell)$ to index the $\ell$-way fold $I_\kk := I_{k_1} \times \cdots \times I_{k_\ell}$ and its complementary product $I_\kk^c := I_{k_1}^c \times \cdots \times I_{k_\ell}^c$ for each $\emph{\textbf{k}} \in [K]^\ell$.
Let 
$$
\hat\eta_\kk = \hat \eta(((W_{\iota, \emph{\textbf{j}}})_{\iota \in [N_\emph{\textbf{j}}]})_{\emph{\textbf{j}} \in I_\kk^c})
$$
be a machine learning estimate of $\eta$ using the subsample $((W_{\iota, \emph{\textbf{j}}})_{\iota \in [N_\emph{\textbf{j}}]})_{\emph{\textbf{j}} \in I_\kk^c}$ for each $\kk \in [K]^\ell$.
Let
\begin{align*}
&\hat J:=\frac{1}{K^\ell}\sum\limits_{\kk\in[K]^\ell}\Enkk\Big[\sumN\psi^a(W_{\imath,\emph{\textbf{j}}};\hat{\eta}_{\kk})\Big] \qquad\text{where}
\\
&\Enkk\Big[\sumN f(W_{\imath,\emph{\textbf{j}}})\Big]:=\frac{1}{|I_\kk|}\sum\limits_{\emph{\textbf{j}}\in I_\kk} \sumN f(W_{\imath,\emph{\textbf{j}} }) \text{ for each } \kk \in [K]^\ell
\end{align*}
for any Borel measurable function $f$, the sum $\sumN$ is treated as zero when $N_\emph{\textbf{j}} = 0$, and
$|I_\kk| := \lfloor\frac{\prod_{i=1}^\ell C_i} {K^\ell}\rfloor$.
With these setup and notations, the multiway DML estimator is defined by
\begin{align}
\tilde \theta =&- \hat J^{-1} \frac{1}{K^\ell}\sumk\Enkk\Big[\sumN\psi^b(W_{\imath,\emph{\textbf{j}}};\hat \eta_{\kk})\Big].
\label{eq:estimator_expression}
\end{align}
Let $|\underline{I_\kk}|=\min\{|I_{k_1}|,...,|I_{k_\ell}|\}$ for a short-hand notation.
Also let $I(\emph{\textbf{j}})$ denote the multiway fold containing the $\emph{\textbf{j}}$-th multiway cluster, i.e., $I(\emph{\textbf{j}}) \subset \times_{i=1}^\ell [C_i]$ satisfies $I_\kk = I(\emph{\textbf{j}})$ for some $\kk \in [K]^\ell$ and $\emph{\textbf{j}} \in I(\emph{\textbf{j}})$.
With these additional notations, we propose to estimate the asymptotic variance of $\sqrt{\underline{C}}(\tilde{\theta}-\theta_0)$ by
\begin{align}
\hat \sigma^2=&\hat J^{-1} \Big[\frac{1}{K^\ell}\sumk
\frac{|\underline{I_\kk}|}{\prodI}
\sum_{i=1}^\ell \sum_{\substack{\emph{\textbf{j}},\emph{\textbf{j}}^{\:\prime}\in I_\kk\\I_i(\emph{\textbf{j}})=I_i(\emph{\textbf{j}}^{\:\prime})}}
\sumN\sumjjj
\psi(W_{\imath,\emph{\textbf{j}}};\tilde \theta,\hat\eta_{\kk}) \psi(W_{\imath,\emph{\textbf{j}}^{\:\prime}};\tilde \theta,\hat\eta_{\kk})'\Big]
(\hat J^{-1})'.
\label{eq:variance_estimator_expression}
\end{align}

\newtheorem*{ex:overview_continued}{Example \ref{ex:overview}, Continued}
\begin{ex:overview_continued}
The two-way DML in Section \ref{sec:multiway_dml} is a special case of the current general methodological framework with $\{I_1^1,...,I_1^K\} = \{I_1,...,I_K\}$, $\{I_2^1,...,I_2^K\} = \{J_1,...,J_K\}$, $\hat\eta_{(k_1,k_2)} = \hat\eta ((W_{j_1 j_2})_{(j_1,j_2) \in ([N] \backslash I_{k_1}) \times ([M] \backslash J_{k_2}})$,
$\hat J=\frac{1}{K^2}\sum_{(k_1,k_2) \in [K]^2} \mathbb{E}_{n,(k_1, k_2)}[\psi^a(W_{j_1 j_2};\hat \eta_{(k_1,k_2)})]$ where 
$\mathbb E_{n,(k_1,k_2)}$ $[f(W_{j_1j_2})] = \frac{1}{|I_{k_1}||J_{k_2}|}\sum_{(j_1,j_2) \in I_{k_1} \times J_{k_2}} f(W_{j_1j_2})$, 
$\tilde \theta =- \hat J^{-1} \frac{1}{K^2}\sum_{(k_1,k_2) \in [K]^2} \mathbb{E}_{n,(k_1,k_2)} \Big[\psi^b(W_{j_1j_2};\hat \eta_{(k_1,k_2)})\Big]$,
and
$
\hat\sigma^2 = \hat J^{-1} \hat\Gamma (\hat J^{-1})'$
where
$\hat \Gamma=$
$
\frac{1}{K^2}\sum_{(k_1,k_2) \in [K]^2}$ \footnotesize
$\Big\{
\frac{|I_{k_1}|\wedge|J_{k_2}|}{(|I_{k_1}||J_{k_2}|)^2}\sum_{j_1\in I_{k_1}}\sum_{j_2,j_2'\in J_{k_2}} \psi(W_{j_1 j_2};\tilde \theta,\hat\eta_{(k_1,k_2)}) \psi(W_{j_1j_2'};\tilde \theta,\hat\eta_{(k_1,k_2)})'
+
\frac{|I_{k_1}|\wedge|J_{k_2}|}{(|I_{k_1}||J_{k_2}|)^2}\sum_{j_1,j_1'\in I_{k_1}}\sum_{j_2\in J_{k_2}} \psi(W_{j_1 j_2};\tilde \theta,\hat\eta_{(k_1,k_2)})\psi(W_{j_1'j_2};\tilde \theta,\hat\eta_{(k_1,k_2)}) '
\Big\}
$.\normalsize
$\triangle$
\end{ex:overview_continued}

We now state assumptions under which (\ref{eq:variance_estimator_expression}) is an asymptotically valid variance estimator for $\sqrt{\underline{C}}(\tilde{\theta}-\theta_0)$ with the multiway DML estimator (\ref{eq:estimator_expression}).
We write $a \lesssim b$ to mean $a \leq cb$ for some $c > 0$ that does not depend on $n$.
We also write $a \lesssim_P b$ to mean $a = O_P(b)$.
For any finite dimensional vector $v$, $\|v\|$ denotes the $\ell_2$ or Euclidean norm of $v$.
For any matrix $A$, $\|A\|$ denotes the induced $\ell_2$-norm of the matrix.
The following assumption concerns the multiway clustered sampling. 

\begin{assumption}[Sampling]
\label{a:sampling 2}
The following conditions hold for each $n$.
\begin{enumerate}[(i)]
\item The array $(N_\emph{\textbf{j}}, (W_{\imath, \emph{\textbf{j}}})_{1\leq \imath\leq \overline{N}})_{\emph{\textbf{j}}\geq \mathbf{1}}$ is an infinite sequence of separately exchangeable random vector. 
That is, for any $\ell$-tuple of permutations $(\pi_1,...,\pi_\ell)$ of $\mathbb N$, we have
\begin{align*}
(N_\emph{\textbf{j}}, (W_{\imath, \emph{\textbf{j}}})_{1\leq \imath \leq \overline{N}})_{\emph{\textbf{j}}\geq \mathbf{1}}\overset{d}{=} (N_{\pi_1(j_1),...,\pi_\ell(j_\ell)}, (W_{\imath, {\pi_1(j_1),...,\pi_\ell(j_\ell)}})_{1\leq \imath \leq \overline{N}})_{\emph{\textbf{j}}\geq \mathbf{1}}.
\end{align*}
\item $(N_\emph{\textbf{j}}, (W_{\imath, \emph{\textbf{j}}})_{1\leq \imath \leq \overline{N}})_{\emph{\textbf{j}}\geq \mathbf{1}}$ is dissociated. 
That is, for any $\textbf{c}\geq \mathbf{1}$, 
$
(N_\emph{\textbf{j}}, (W_{\imath, \emph{\textbf{j}}})_{1\leq \imath \leq \overline{N}})_{\mathbf{1} \leq \emph{\textbf{j}}\leq \emph{\textbf{c}}}
$
is independent of 
$
(N_{\emph{\textbf{j}}^{\:\prime}}, (W_{\imath', \emph{\textbf{j}}^{\:\prime}})_{1\leq\imath'\leq \overline{N}})_{\emph{\textbf{j}}^{\:\prime}\geq \emph{\textbf{c}}+\mathbf{1}}
$
\item $ E(N_\mathbf{1})>0$ and $N_{\emph{\textbf{j}}}\leq \overline{N}$ for each $\textbf{1}\leq \emph{\textbf{j}}\leq \emph{\textbf{C}}$, where $\overline{N} \in \mathbb{N}$ does not depend on $n$.
\item The econometrician observes 
$
(N_\emph{\textbf{j}}, (W_{\imath, \emph{\textbf{j}}})_{1\leq \imath \leq N_{\emph{\textbf{j}}}})_{\mathbf{1} \leq \emph{\textbf{j}}\leq \emph{\textbf{C}}}
$. 
\end{enumerate}
\end{assumption}
\begin{remark}The dependence among $\left(W_{\imath,\emph{\textbf{j}}}\right)_{\imath \geq 1}$ in each cell $\emph{\textbf{j}}$ is left unrestricted in this assumption. 
Assumption \ref{a:sampling 2} is similar to Assumption 1 of \cite{DDG18}, except for $\overline{N}$. We introduce $\overline{N}$ to simplify some concentration arguments.
\end{remark}

Let $c_0>0$, $c_1>0$, $s>0$, $q\ge 4$ 
be some finite constants with $c_0\le c_1$. 
Let $\{\delta_n\}_{n\ge 1}$ (estimation errors) and $\{\Delta_n\}_{n\ge 1}$ (probability bounds) be sequences of positive constants that converge to zero such that $\delta_n \ge \C^{-1/2}$. 
Let $K\ge 2$ be a fixed integer. 
Let $(N_\textbf{0},(W_{\imath,\textbf{0}})_{0\leq \imath \leq \overline{N}})$  denote an independent copy of $(N_\textbf{1},(W_{\imath,\textbf{1}})_{1\leq \imath \leq \overline{N}})$ and therefore is independent from the data and the random set $\mathcal T_n$ of nuisance realization. 
With these notations, we state the following assumptions for the model.
\begin{assumption}[Linear Neyman Orthogonal Score]\label{a:linear_orthogonal_score 2}
For all $\C\ge 3$ and $P\in \mathcal P_n$, the following conditions hold.
\begin{enumerate}[(i)]
\item The true parameter value $\theta_0$ satisfies (\ref{eq:existance_condition 2}).
\item $\psi$ is linear in the sense that it satisfies (\ref{eq:linear_score2}).
\item The map $\eta \mapsto \Ep\Big[\sumNN\psi(W_{\imath,\textbf{0}};\theta,\eta)\Big]$ is twice continuously Gateaux differentiable on $T$.
\item $\psi$ satisfies the Neyman near orthogonality condition at $(\theta_0,\eta_0)$ as
\begin{align*}
\lambda_n:=\sup_{\eta \in \mathcal T_n}\Big\| \partial_\eta \Ep\Big[\sumNN\psi(W_{\imath,\textbf{0}};\theta_0,\eta_0)[\eta-\eta_0]\Big] \Big\|\le \delta_n \C^{-1/2}.
\end{align*}
\item The identification condition holds as the singular values of the matrix $J_0:=\Ep\Big[\sumNN\psi^a(W_{\imath,\textbf{0}};\eta_0)\Big]$ are between $c_0$ and $c_1$.
\end{enumerate}
\end{assumption}
\begin{assumption}[Score Regularity and Nuisance Parameter Estimators]\label{a:regularity_nuisance_parameters 2}
For all $\C\ge 3$ and $P\in \mathcal P_n$, the following conditions hold.
\begin{enumerate}[(i)]
\item The realization set $\mathcal T_n$ contains $\eta_0$, and the nuisance parameter estimator $\hat \eta_\kk = \hat\eta((W_{\iota, \emph{\textbf{j}}})_{\iota \in [N_\emph{\textbf{j}}]})_{\emph{\textbf{j}} \in I_\kk^c}$ belongs to the realization set $\mathcal T_n$ for each $\kk \in [K]^\ell$ with probability at least $1-\Delta_n$.
\item The following moment conditions hold:
\begin{align*}
m_n:=& \sup_{\eta\in \T_n}(\Ep\Big[\Big\|\sumNN\psi(W_{\imath,\textbf{0}};\theta_0,\eta)\Big\|^q\Big])^{1/q} \le c_1,\\
m_n':=& \sup_{\eta\in \T_n}(\Ep\Big[\Big\|\sumNN\psi^a(W_{\imath,\textbf{0}};\eta)\Big\|^q\Big])^{1/q} \le c_1.
\end{align*}
\item The following conditions on the rates $r_n$, $r_n'$ and $\lambda_n'$ hold:
\begin{align*}
r_n:=& \sup_{\eta\in \T_n}
\Big\|\Ep\Big[\sumNN\psi^a(W_{\imath,\textbf{0}};\eta)\Big]-\Ep\Big[\sumNN\psi^a(W_{\imath,\textbf{0}};\eta_0)\Big]\Big\|\le \delta_n,\\
r_n':=& \sup_{\eta\in \T_n}
\Big(\Big\|\Ep\Big[\sumNN\psi(W_{\imath,\textbf{0}};\theta_0,\eta)\Big]-\Ep\Big[\sumNN\psi(W_{\imath,\textbf{0}};\theta_0,\eta_0)\Big]\Big\|^2\Big)^{1/2}\le \delta_n,\\
\lambda_n'= & \sup_{r\in (0,1),\eta\in \T_n}\Big\|\partial^2_r \Ep\Big[\sumNN\psi (W_{\imath,\textbf{0}};\theta_0,\eta_0+r(\eta-\eta_0)) \Big]\Big \|\le \delta_n/\sqrt{\C}.
\end{align*}
\item All eigenvalues of the matrix 
\begin{align*}
\Gamma:=\sum_{i=1}^\ell\bar\mu_i \Gamma_i =\sum_{i=1}^\ell\bar\mu_i\Ep \left[\sum_{\imath=1}^{N_\textbf{1}}\sum_{\imath'=1}^{N_{2_{\textbf{i}}}}\psi(W_{\imath,\textbf{1}};\theta_0,\eta_0)\psi(W_{\imath',2_i};\theta_0,\eta_0)'\right] 
\end{align*}
are bounded from below by $c_0$, where $2_i$ denotes the $\ell-$tuple vector with 2 in each entry but for 1 in the $i$-th entry.
\end{enumerate}
\end{assumption}

The following theorems generalize Theorems \ref{theorem:main_result_linear} and \ref{theorem:variance_estimator_linear} to cover general $\ell$-way cluster sampling.
Their proofs are contained in Section \ref{sec:proofs_for_general_multiway}.
\begin{theorem}[Main Result]\label{theorem:main_result_linear 2}
Suppose that Assumptions \ref{a:sampling 2}, \ref{a:linear_orthogonal_score 2} and \ref{a:regularity_nuisance_parameters 2} are satisfied.
If $\delta_n\ge \C^{-1/2}$ for all $\C\ge 1$, then
\begin{align*}
\sqrt{\C}\sigma^{-1}(\tilde \theta - \theta_0)=\frac{\sqrt{\C}}{\prod_C}\sumjj\sumN
\bar \psi(W_{\imath,\emph{\textbf{j}}})+\Op(\rho_n)\leadsto N(0,I_{d_\theta})
\end{align*}
holds uniformly for all $P\in\mathcal P_n$, where 
$\prod_C=\prod\limits_{i=1}^\ell C_i$,
the influence function takes the form $\bar \psi(\cdot):=-\sigma^{-1}J_0^{-1} \psi(\cdot;\theta_0,\eta_0)$,
the size of the remainder terms follows
\begin{align*}
\rho_n :=\C^{-1/2} + r_n +r_n' + \C^{1/2} \lambda_n + \C^{1/2} \lambda_n'\lesssim \delta_n,
\end{align*}
and the asymptotic variance is given by
\begin{align}
\sigma^2:=J_0^{-1}\Gamma (J_0^{-1})'. \label{eq:population_variance}
\end{align}
\end{theorem}

\begin{theorem}[Variance Estimator]\label{theorem:variance_estimator_linear 2}
Under the assumptions required by Theorem \ref{theorem:main_result_linear 2}, we have
\begin{align*}
\hat \sigma^2=\sigma^2 +\Op(\rho_n).
\end{align*}
Furthermore, the statement of Theorem \ref{theorem:main_result_linear 2} holds true with $\hat \sigma^2$ in place of $\sigma^2$.
\end{theorem}

%
%

\section{Proofs of the Extended Results}\label{sec:proofs_for_general_multiway}

\subsection{Proof of Theorem \ref{theorem:main_result_linear 2}}\label{sec:theorem:main_result_linear 2}
\begin{proof}
Let $\E_n$ denote the event $\hat \eta_{(k_1,...,k_\ell)} \in \T_n$ for all $\left(k_1,...k_\ell\right)\in [K]^\ell$ and define $\emph{\textbf{k}}:=(k_1,...,k_\ell)$. Assumption \ref{a:regularity_nuisance_parameters 2} (i) implies $P_n(\E_n)\ge 1- K^\ell\Delta_n$.
Let  $\textbf{e}\in \{0,1\}^\ell$, and define $\mathcal{A}_{\textbf{e}}:=\{(\emph{\textbf{j}},\emph{\textbf{j}}^{\:\prime}):\textbf{1}\leq \emph{\textbf{j}},\emph{\textbf{j}}^{\:\prime} \leq \textbf{C}: \forall i=1,...,\ell, e_i=1\Leftrightarrow j_i=j_i'\}$, and $\boldsymbol{\varepsilon}_m:=\{\textbf{e}\in\{0,1\}^\ell:\sum_{i'=1}^\ell e_{i'}=m\}$.\\
\noindent \textbf{Step 1.}
This is the main step showing linear representation and asymptotic normality for the proposed estimator. 
Denote 
\begin{align*}
&\hat J:=\frac{1}{K^\ell}\sum\limits_{\kk\in[K]^\ell}\Enkk\Big[\sumN\psi^a(W_{\imath,\emph{\textbf{j}}};\hat{\eta}_{\kk})\Big],
\qquad R_{n,1}:=\hat J - J_0,\\
&R_{n,2}:=\frac{1}{K^\ell} \sum\limits_{\kk\in[K]^\ell}\Enkk\Big[\sumN\psi(W_{\imath,\emph{\textbf{j}}};\theta_0,\hat{\eta}_{\kk})\Big]-\frac{1}{\prod_C}\sumjj\sumN\psi(W_{\imath,\emph{\textbf{j}}};\theta_0,\eta_0).
\end{align*}
We will later show in Steps 2, 3, 4 and 5, respectively, that
\begin{align}
&\|R_{n,1}\|=\Opn(\C^{-1/2} + r_n), \label{eq:step2}\\
&\|R_{n,2}\|=\Opn(\C^{-1/2}r_n'+ \lambda_n +  \lambda_n'), \label{eq:step3}\\
&\Big\|\sqrt{\C}\frac{1}{\prod_C}\sumjj\sumN\psi(W_{\imath,\emph{\textbf{j}}};\theta_0,\eta_0)\Big\|=\Opn(1), \label{eq:step4}\\
&\|\sigma^{-1}\|=\Opn(1). \label{eq:step5}
\end{align}
Then,
under Assumptions \ref{a:linear_orthogonal_score 2} and \ref{a:regularity_nuisance_parameters 2}, $\C^{-1/2}+r_N\le \rho_n=o(1)$ and all singular values of $J_0$ are bounded away from zero.
Therefore, with $P_n$-probability at least $1-o(1)$, all singular values of  $\hat J$ are bounded away from zero. Thus with the same $P_n$ probability, the multiway DML solution is uniquely written as
\begin{align*}
\tilde \theta =- \hat J^{-1} \frac{1}{K^\ell}\sumk\Enkk\Big[\sumN\psi^b(W_{\imath,\emph{\textbf{j}}};\hat \eta_{\kk})\Big],
\end{align*}
and 
\begin{align}
\sqrt{\C}(\tilde \theta - \theta_0)
=&
-\sqrt{\C} \hat J^{-1}\frac{1}{K^\ell} \sumk\Big(\Enkk\Big[\sumN\psi^b(W_{\imath,\emph{\textbf{j}}};\hat \eta_{\kk})\Big] +\hat J \theta_0\Big) \nonumber\\
=&-\sqrt{\C} \hat J^{-1}\frac{1}{K^\ell} \sumk \Enkk\Big[\sumN\psi(W_{\imath,\emph{\textbf{j}}};\theta_0,\hat \eta_{\kk})\Big] \nonumber\\
=&
-\Big(J_0 + R_{n,1}\Big)^{-1} \times \Big(
\frac{\sqrt{\C}}{\prod_C} \sumjj\sumN \psi(W_{\imath,\emph{\textbf{j}}};\theta_0,\eta_0) +\sqrt{\C} R_{n,2}
\Big).
\label{eq:step16}
\end{align}
Using the fact that 
\begin{align*}
\Big(J_0 + R_{n,1}\Big)^{-1}  - J_0^{-1}=-(J_0 +R_{n,1})^{-1}R_{n,1} J_0^{-1},
\end{align*}
we have 
\begin{align*}
\|(J_0+R_{n,1})^{-1} - J_0^{-1}\| = &
\|(J_0 +R_{n,1})^{-1}R_{n,1} J_0^{-1}\|\le  \|(J_0 +R_{n,1})^{-1}\|\,\|R_{n,1}\|\, \|J_0^{-1}\|\\
=&\Opn(1)\Opn(\C^{-1/2}+ r_n)\Opn(1)=\Opn(\C^{-1/2}+ r_n).
\end{align*}
Furthermore, $r_n'+\sqrt{\C}(\lambda_n+\lambda_n')\le \rho_n=o(1)$, it holds that 
\begin{align*}
\Big\|\frac{\sqrt{\C}}{\prod_C} \sumjj \sumN\psi(W_{\imath,\emph{\textbf{j}}};\theta_0,\eta_0) +\sqrt{\C} R_{n,2}\Big\|
\le&
\Big\|\frac{\sqrt{\C}}{\prod_C} \sumjj\sumN \psi(W_{\imath,\emph{\textbf{j}}};\theta_0,\eta_0)\Big\| +\Big\|\sqrt{\C} R_{n,2}\Big\|  \\
=&
\Opn(1) +\opn(1)=\Opn(1),
\end{align*}
where the first equality is due to (\ref{eq:step4}) and (\ref{eq:step5}).
Combining above two bounds gives 
\begin{align}
\Big\|\Big(J_0 + R_{n,1}\Big)^{-1}  - J_0^{-1}\Big\|\times\Big\|\frac{\sqrt{\C}}{\prod_C} \sumjj \sumN\psi(W_{\imath,\emph{\textbf{j}}};\theta_0,\eta_0) +\sqrt{\C} R_{n,2}\Big\| =&
\Opn(\C^{-1/2}+ r_n)\Opn(1) \nonumber\\
=&
\Opn(\C^{-1/2}+ r_n).
\label{eq:step1_2}
\end{align}
Therefore, from (\ref{eq:step5}), (\ref{eq:step16}) and (\ref{eq:step1_2}), we have
\begin{align*}
\sqrt{\C}\sigma^{-1}(\tilde \theta - \theta_0)=&
\frac{\sqrt{\C}}{\prod_C} \sumjj \sumN\bar\psi(W_{\imath,\emph{\textbf{j}}}) +\Opn(\rho_n) .
\end{align*}
The first term on the RHS above can be written as $\GC \bar\psi$. 
Applying Lemma \ref{lemma:hajek_multiway}, we obtain the independent linear representation
\begin{align*}
H_n \bar\psi:=\sum_{j_1=1}^{C_1} \frac{\sqrt{\C}}{C_1} \Epn\Big[\sumN\bar \psi (W_{\imath,\emph{\textbf{j}}})\Big|U_{j_1,0...0}\Big] +...+ \sum_{j_\ell=1}^{C_\ell} \frac{\sqrt{\C}}{C_\ell} \Epn\Big[\sumN\bar \psi (W_{\imath,\emph{\textbf{j}}})\Big|U_{0...0,j_\ell}\Big]
\end{align*}
and it holds $P_n$-a.s. that
\begin{align*}
V_n(\GC \bar\psi)=& V_n(H_n \bar\psi) + O(\C^{-1})=J_0^{-1}\Gamma (J_0^{-1})' + O(\C^{-1})
\qquad\text{and}\\
\GC \bar\psi =& H_n \bar\psi + \Op(\C^{-1/2}),
\end{align*}
where $V_n(\cdot)=\Epn[(\cdot-\Epn[\cdot])^2]$.
Under Assumption \ref{a:regularity_nuisance_parameters 2} (iv).
Recall that $q\ge 4$, the third moments of both summands of $H_n \bar\psi$ are bounded over $n$ under Assumptions \ref{a:linear_orthogonal_score 2}(v) and \ref{a:regularity_nuisance_parameters 2} (ii) (iv).
We have verified all the conditions for Lyapunov's CLT. An application of Lyapunov's CLT and Cramer-Wold device gives
\begin{align*}
&H_n \bar\psi \leadsto N(0,I_{d_\theta})
\end{align*}
and an application of Theorem 2.7 of \cite{vdV98} concludes the proof.

\noindent \textbf{Step 2.}
Since $K$ is fixed, it suffices to show for any $\kk\in [K]^\ell$,
\begin{align*}
\Big\|\Enkk\Big[\sumN\psi^a(W_{\imath,\emph{\textbf{j}}};\hat \eta_{\kk})\Big]-\Ep\Big[\sumNN\psi^a(W_{\imath,\textbf{0}};\eta_0)\Big]\Big\|
=
\Opn(\C^{-1/2} + r_n).
\end{align*}
Fix $\kk\in [K]^\ell$,
\begin{align*}
&\Big\|\Enkk\Big[\sumN\psi^a(W_{\imath,\emph{\textbf{j}}};\hat \eta_{\kk})\Big]-\Epn\Big[\sumNN\psi^a(W_{\imath,\textbf{0}};\eta_0)\Big]\Big\|
\le \mathcal I_{1,\kk} + \mathcal I_{2,\kk},
\end{align*}
where
\begin{align*}
\mathcal I_{1,\kk} &:= \Big\|\Enk\Big[\sumN\psi^a(W_{\imath,\emph{\textbf{j}}};\hat \eta_{\kk})\Big]-\Epn\Big[\sumN\psi^a(W_{\imath,\emph{\textbf{j}}};\hat \eta_{\kk})\Big|I_{k_1}^c \times...\times I_{k_\ell}^c\Big]\Big\|
,\\
\mathcal I_{2,\kk} &:= \Big\|\Epn\Big[\sumN\psi^a(W_{\imath,\emph{\textbf{j}}};\hat \eta_{\kk})\Big|I_{k_1}^c \times...\times I_{k_\ell}^c\Big]-\Epn\Big[\sumNN\psi^a(W_{\imath,\textbf{0}};\eta_0)\Big]\Big\|.
\end{align*}
Notice that $\mathcal I_{2,\kk}\le r_n$ with $P_n$-probability $1-o(1)$ follows directly from Assumptions \ref{a:sampling 2} (ii) and \ref{a:regularity_nuisance_parameters 2} (iii). 
Now denote $\tilde\psi^a_{\emph{\textbf{j}},m}=\sumN\psi^a_m(W_{\imath,\emph{\textbf{j}}};\hat \eta_{\kk})-\Epn\Big[\sumN\psi^a_m(W_{\imath,\emph{\textbf{j}}};\hat \eta_{\kk})\Big|I_{k_1}^c\times...\times I_{k_\ell}^c\Big]$ and $\tilde \psi^a_{\emph{\textbf{j}}}=(\tilde \psi^a_{\emph{\textbf{j}},m})_{m\in [d_\theta]}$, and ${|\underline{I_\kk}|}=\min\{|I_{k_1}|,...,|I_{k_\ell}|\}$. 
Let us denote $I_\kk:=\left(I_{k_1}\times...\times I_{k_\ell}\right)$ and $I_{\kk}^c:=(I_{k_1}^c\times...\times I_{k_\ell}^c)$.
Let $\emph{\textbf{j}}\mapsto I(\emph{\textbf{j}}) \in \mathcal{I}$, and define $\mathcal{B}_{\textbf{e}}:=\{(\emph{\textbf{j}}, \emph{\textbf{j}}^{\:\prime}):\forall i=1,...,\ell, e_i=1\Leftrightarrow I_i(\emph{\textbf{j}})=I_i(\emph{\textbf{j}}^{\:\prime}): \emph{\textbf{j}},\emph{\textbf{j}}^{\:\prime}\in \mathcal{I}\}$,
where $\mathcal{I}:=\{I_1^1,...,I_1^K\}\times...\times\{I_\ell^1,...,I_\ell^K\}$, and $\boldsymbol{\epsilon}_m:=\{\textbf{e}\in\{0,1\}^\ell:\sum_{i'=1}^\ell e_{i'}=m\}$.
To bound $\mathcal I_{1,\kk}$, note that conditional on $I^c_{\kk}$, it holds that
\begin{align*}
\Epn[\mathcal I_{1,\kk}^2|I^c_{\kk}]
=&
\Epn\Big[\Big\|\Enkk\Big[\sumN\psi^a(W_{\imath,\emph{\textbf{j}}};\hat \eta_{\kk})\Big]-\Epn\Big[\sumN\psi^a(W_{\imath,\emph{\textbf{j}}};\hat \eta_{\kk})\Big|I^c_{\kk}\Big]\Big\|^2\Big|I^c_{\kk}\Big]\\
=&
\frac{1}{\prodI}\Epn\Big[\sum_{m=1}^{d_\theta}\Big(\sum\limits_{\emph{\textbf{j}} \in I_\kk} \tilde\psi^a_{\emph{\textbf{j}},m}\Big)^2\Big|I^c_{\kk}\Big]\\
=&
\frac{1}{\prodI}\sum_{\textbf{e}\in\boldsymbol{\epsilon}_1}\sum_{\left(\emph{\textbf{j}}^{\:\prime},\emph{\textbf{j}}\right)\in \mathcal{B}_\textbf{e}}\Epn\Big[\sum_{m=1}^{d_\theta}\tilde \psi^a_{\emph{\textbf{j}},m}\tilde \psi^a_{\emph{\textbf{j}}^{\:\prime},m}\Big|I^c_{\kk}\Big]\\
&+
\frac{1}{\prodI}\sum_{r=2}^\ell \sum_{\textbf{e}\in\boldsymbol{\epsilon}_r}\sum_{\left(\emph{\textbf{j}}^{\:\prime},\emph{\textbf{j}}\right)\in \mathcal{B}_\textbf{e}}\Epn\Big[\sum_{m=1}^{d_\theta}\tilde \psi^a_{\emph{\textbf{j}},m}\tilde \psi^a_{\emph{\textbf{j}}^{\:\prime},m}\Big|I^c_{\kk}\Big]
\\
&+
\frac{1}{\prodI}\sum_{\textbf{e}\in\boldsymbol{\epsilon}_0}\sum_{\left(\emph{\textbf{j}}^{\:\prime},\emph{\textbf{j}}\right)\in \mathcal{B}_\textbf{e}}\Epn\Big[\sum_{m=1}^{d_\theta}\tilde \psi^a_{\emph{\textbf{j}},m}\tilde \psi^a_{\emph{\textbf{j}}^{\:\prime},m}\Big|I^c_{\kk}\Big]
\\
=&\frac{1}{\prodI}\sum_{\textbf{e}\in\boldsymbol{\epsilon}_1}\sum_{\left(\emph{\textbf{j}}^{\:\prime},\emph{\textbf{j}}\right)\in \mathcal{B}_\textbf{e}}\Epn[\langle\tilde\psi^a_{\emph{\textbf{j}}},\tilde\psi^a_{\emph{\textbf{j}}^{\:\prime}}\rangle|I^c_{\kk}]+R+0
\\
\lesssim&
\frac{1}{|\underline{I_\kk}|}\Epn\Big[\Big\|\sumN\psi^a(W_{\imath,\emph{\textbf{j}}};\hat\eta_{\kk})-\Epn\Big[\sumN\psi^a(W_{\imath,\emph{\textbf{j}}};\hat\eta_{\kk})\Big|I^c_{\kk}\Big]\Big\|^2\Big|I^c_{\kk}\Big]
\\
\leq&\frac{1}{|\underline{I_\kk}|}\Epn\Big[\Big\|\sumN\psi^a(W_{\imath,\emph{\textbf{j}}};\hat\eta_{\kk})\Big\|^2\Big|I^c_{\kk}\Big]
\leq
\frac{c_1^2}{|\underline{I_\kk}|}.
\end{align*}
\noindent In the third equility, the last term corresponds to the covariance between cells sharing no common cluster. By independence, the last term is zero. Let us  denote the second term in the third equality by $R$.  Under Cauchy-Schwarz inequality and Assumption \ref{a:sampling 2} (ii), 
\begin{align}
|R|\leq \frac{1}{\prodI}\sum_{\textbf{e}\in\cup_{l=2}^\ell \boldsymbol{\epsilon}_l}|\mathcal{B}_\textbf{e}|\Epn\Big[\sum_{m=1}^{d_\theta}(\tilde\psi^a_{\emph{\textbf{j}},m})^2\Big|I^c_{\kk}\Big].
\end{align}
For $r\geq 1$ and $\textbf{e}\in \boldsymbol{\epsilon}_r$, we have 
\begin{align}
|\mathcal{B}_\textbf{e}|=|I_\kk|\times\prod_{i:e_i=0}(|I_{k_i}|-1).
\end{align}
Therefore, $R=O(\underline{|I_\kk|}^{-2})$. Note that $\C\lesssim |\underline{I_\kk}|\lesssim\C$. Hence an application of Lemma \ref{lemma:conditional_convergence} (i) implies 
$
\mathcal I_{1,\kk}=\Opn(\C^{-1/2}).
$
This completes a proof of (\ref{eq:step2}).

\noindent \textbf{Step 3.}
It again suffices to show that for any $\kk\in [K]^\ell$, one has
\begin{align*}
\Big\|\Enk\Big[\sumN\psi(W_{\imath,\emph{\textbf{j}}};\theta_0,\hat\eta_{\kk})\Big]
- \frac{1}{\prodll}\sum_{\emph{\textbf{j}}\in I_\kk}\sumN\psi(W_{\imath,\emph{\textbf{j}}};\theta_0,\eta_0)\Big\|=\Opn(\C^{-1/2}r_n'+ \lambda_n +  \lambda_n').
\end{align*}
Denote 
$$\Gnkk\Big[\sumN\phi(W_{\imath,\emph{\textbf{j}}})\Big]=\frac{\sqrt{\C}}{\prodll}\sum_{\emph{\textbf{j}}\in I_\kk}\sumN\Big( \phi(W_{\imath,\emph{\textbf{j}}}) - \int \phi(w) dP_n \Big). $$
Then 
\begin{align*}
&\Big\|\Enk\Big[\sumN\psi(W_{\imath,\emph{\textbf{j}}};\theta_0,\hat\eta_{\kk})\Big]
- \frac{1}{\prodll} \sum_{\emph{\textbf{j}}\in I_\kk}\sumN\psi(W_{\imath,\emph{\textbf{j}}};\theta_0,\eta_0)\Big\|
\le
\frac{\mathcal I_{3,\kk}+\mathcal I_{4,\kk}}{\sqrt{\C}},
\end{align*}
where
\begin{align*}
\mathcal I_{3,\kk}
:=&
\Big\|\Gnkk\Big[\sumN\psi(W_{\imath,\emph{\textbf{j}}};\theta_0,\hat \eta_{\kk})\Big]-\Gnkk\Big[\sumN\psi(W_{\imath,\emph{\textbf{j}}};\theta_0,\eta_0)\Big]\Big\|,
\\
\mathcal I_{4,\kk}
:=&\sqrt{\C}
\Big\|
\Epn\Big[\sumN\psi(W_{\imath,\emph{\textbf{j}}};\theta_0,\hat \eta_{\kk})\Big|I_\kk\Big]- \Epn\Big[\sumNN\psi(W_{\imath,\textbf{0}};\theta_0,\eta_0)\Big]
\Big\|.
\end{align*}
Denote $\tilde \psi_{\emph{\textbf{j}},m}:=\sumN\psi_m(W_{\imath,\emph{\textbf{j}}};\theta_0,\hat \eta_{\kk})-\sumN\psi_m(W_{\imath,\emph{\textbf{j}}};\theta_0,\eta_0)$ and $\tilde \psi_{\emph{\textbf{j}}}=(\tilde \psi_{\emph{\textbf{j}},m})_{m\in[d_\theta]}$. To bound $\mathcal I_{3,\kk}$, notice that using a similar argument as for the bound of $\mathcal I_{1,\kk}$, one has
{\small
\begin{align*}
\Epn[\|\mathcal I_{3,\kk}\|^2|I^c_{\kk}]
=&
\Epn\Big[\Big\|\Gnkk\Big[\sumN\psi(W_{\imath,\emph{\textbf{j}}};\theta_0,\hat \eta_{\kk})-\sumN\psi(W_{\imath,\emph{\textbf{j}}};\theta_0,\eta_0)\Big]\Big\|^2\Big|I^c_{\kk}\Big]\\
=&
\Epn\Big[\frac{\C}{\prodI}\sum_{m=1}^{d_\theta}\Big\{\sum_{\emph{\textbf{j}}\in I_\kk} \Big(\tilde\psi_{\emph{\textbf{j}},m}
-\Epn\tilde\psi_{\emph{\textbf{j}},m}
\Big)\Big\}^2\Big|I^c_{\kk}\Big]\\
=&
\frac{\C}{\prodI}\sum_{\textbf{e}\in\boldsymbol{\epsilon}_1}\sum_{\left(\emph{\textbf{j}},\emph{\textbf{j}}^{\:\prime}\right)\in \mathcal{B}_\textbf{e}}\Epn\Big[\sum_{m=1}^{d_\theta}\Big(\tilde\psi_{\emph{\textbf{j}},m}-\Epn\tilde\psi_{\emph{\textbf{j}},m}
\Big)\Big(\tilde\psi_{\emph{\textbf{j}}^{\:\prime},m}-\Epn\tilde\psi_{\emph{\textbf{j}}^{\:\prime},m}
\Big)\Big|I^c_{\kk}\Big]\\
&+
\frac{\C}{\prodI}\sum_{r=2}^\ell\sum_{\textbf{e}\in\boldsymbol{\epsilon}_r}\sum_{\left(\emph{\textbf{j}},\emph{\textbf{j}}^{\:\prime}\right)\in \mathcal{B}_\textbf{e}} \Epn\Big[\sum_{m=1}^{d_\theta}\Big(\tilde\psi_{\emph{\textbf{j}},m}-\Epn\tilde\psi_{\emph{\textbf{j}},m}
\Big)\Big(\tilde\psi_{\emph{\textbf{j}}^{\:\prime},m}-\Epn\tilde\psi_{\emph{\textbf{j}}^{\:\prime},m}
\Big)\Big|I^c_{\kk}\Big]\\
&+
\frac{\C}{\prodI}\sum_{\textbf{e}\in\boldsymbol{\epsilon}_0}\sum_{\left(\emph{\textbf{j}},\emph{\textbf{j}}^{\:\prime}\right)\in \mathcal{B}_\textbf{e}}\Epn\Big[\sum_{m=1}^{d_\theta}\Big(\tilde\psi_{\emph{\textbf{j}},m}-\Epn\tilde\psi_{\emph{\textbf{j}},m}
\Big)\Big(\tilde\psi_{\emph{\textbf{j}}^{\:\prime},m}-\Epn\tilde\psi_{\emph{\textbf{j}}^{\:\prime},m}
\Big)\Big|I^c_{\kk}\Big]
\\
=&
\frac{\C}{\prodI}\sum_{\textbf{e}\in\boldsymbol{\epsilon}_1}\sum_{\left(\emph{\textbf{j}},\emph{\textbf{j}}^{\:\prime}\right)\in \mathcal{B}_\textbf{e}} \Epn\Big[\langle\tilde\psi_{\emph{\textbf{j}}}-\Epn\tilde\psi_{\emph{\textbf{j}}}
,\tilde\psi_{\emph{\textbf{j}}^{\:\prime}}-\Epn\tilde\psi_{\emph{\textbf{j}}^{\:\prime}}
\rangle\Big|I^c_{\kk}\Big]+R'+0\\
\lesssim&
 \Epn\Big[\Big\|\sumN\psi(W_{\imath,\emph{\textbf{j}}};\theta_0,\hat\eta)-\sumN\psi(W_{\imath,\emph{\textbf{j}}};\theta_0,\eta_0)-\Epn\Big[\sumN\psi(W_{\imath,\emph{\textbf{j}}};\theta_0,\hat\eta)-\sumN\psi(W_{\imath,\emph{\textbf{j}}};\theta_0,\eta_0)\Big]\Big\|^2\Big|I^c_{\kk}\Big] \\
\leq&
 \Epn\Big[\Big\|\sumN\psi(W_{\imath,\emph{\textbf{j}}};\theta_0,\hat\eta)-\sumN\psi(W_{\imath,\emph{\textbf{j}}};\theta_0,\eta_0)\Big\|^2\Big|I^c_{\kk}\Big] \\
 \le&
 \sup_{\eta \in \T_n}\Epn\Big[\Big\|\sumNN\psi(W_{\imath,\textbf{0}};\theta_0,\eta)-\sumNN\psi(W_{\imath,\textbf{0}};\theta_0,\eta_0)\Big\|^2\Big|I^c_{\kk}\Big] \\
 =&
  \sup_{\eta \in \T_n}\Epn\Big[\Big\|\sumNN\psi(W_{\imath,\textbf{0}};\theta_0,\eta)-\sumNN\psi(W_{\imath,\textbf{0}};\theta_0,\eta_0)\Big\|^2\Big] = (r_n')^2,
\end{align*}}
where the first inequality follows from Cauchy-Schwartz's inequality, the second-to-last equality is due to Assumption \ref{a:sampling 2}, and the last equality is due to Assumption \ref{a:regularity_nuisance_parameters 2} (iii). Using the similar argument for $R$, we have $R'=O(\underline{C}^{-1})$.

Hence, $\mathcal I_{3,\kk}=\Opn(r_n')$.
To bound $\mathcal I_{4,\kk}$, let
\begin{align*}
f_{\kk}(r):=
\Epn\Big[\sumN\psi(W_{\imath,\emph{\textbf{j}}};\theta_0,\eta_0 + r(\hat \eta_{\kk}-\eta_0))\Big|I^c_{\kk}\Big]
-
\Epn\Big[\sumNN\psi(W_{\imath,\textbf{0}};\theta_0,\eta_0)\Big],
\qquad r\in[0,1].
\end{align*}
An application of the mean value expansion coordinate-wise gives
\begin{align*}
f_{\kk}(1)=f_{\kk}(0)+f_{\kk}'(0)+f_{\kk}''(\tilde r)/2,
\end{align*}
where $\tilde r \in (0,1)$.
Note that $f_{\kk}(0)=0$ under Assumption \ref{a:linear_orthogonal_score 2} (i), and
\begin{align*}
\|f_{\kk}'(0)\|=\Big\| \partial_\eta \Epn \Big[\sumN\psi(W;\theta_0,\eta_0)[\hat \eta_{\kk}-\eta_0]\Big] \Big\|\le \lambda_n
\end{align*}
under Assumption \ref{a:linear_orthogonal_score 2} (iv).
Moreover, under Assumption \ref{a:regularity_nuisance_parameters 2} (iii), on the event $\E_n$, we have
\begin{align*}
\|f_{\kk}''(\tilde r)\|\le \sup_{r\in(0,1)}\|f_{\kk}''(r)\|\le \lambda_n'.
\end{align*}
This completes a proof of (\ref{eq:step3}).

\noindent \textbf{Step 4.}
Note that
\begin{align*}
&\Epn\Big[\Big\|\frac{\sqrt{\C}}{\prod_C}\sumjj\sumN\psi(W_{\imath,\emph{\textbf{j}}};\theta_0,\eta_0)\Big\|^2\Big]
\\
&=
\frac{\C}{\prod_C^2}\Epn\Big[\sum_{m=1}^{d_\theta}\Big(\sumjj\sumN\psi_m(W_{\imath,\emph{\textbf{j}}};\theta_0,\eta_0)\Big)^2\Big]\\
=&
\frac{\C}{\prod_C^2}\sum_{\textbf{e}\in\boldsymbol{\varepsilon}_1}\sum_{\left(\emph{\textbf{j}},\emph{\textbf{j}}^{\:\prime}\right)\in \mathcal{A}_\textbf{e}}\Epn\Big[\sum_{m=1}^{d_\theta}\sumN\sumjjj\psi_m(W_{\imath,\emph{\textbf{j}}};\theta_0,\eta_0)\psi_m(W_{\imath',\emph{\textbf{j}}^{\:\prime}};\theta_0,\eta_0)\Big]\\
&+
\frac{\C}{\prod_C^2}\sum_{r=2}^\ell\sum_{\textbf{e}\in\boldsymbol{\varepsilon}_r}\sum_{\left(\emph{\textbf{j}},\emph{\textbf{j}}^{\:\prime}\right)\in \mathcal{A}_\textbf{e}}\Epn\Big[\sum_{m=1}^{d_\theta}\sumN\sumjjj\psi_m(W_{\imath,\emph{\textbf{j}}};\theta_0,\eta_0)\psi_m(W_{\imath',\emph{\textbf{j}}^{\:\prime}};\theta_0,\eta_0)\Big]\\
&+
\frac{\C}{\prod_C^2}\sum_{\textbf{e}\in\boldsymbol{\varepsilon}_0}\sum_{\left(\emph{\textbf{j}},\emph{\textbf{j}}^{\:\prime}\right)\in \mathcal{A}_\textbf{e}}\Epn\Big[\sum_{m=1}^{d_\theta}\sumN\sumjjj\psi_m(W_{\imath,\emph{\textbf{j}}};\theta_0,\eta_0)\psi_m(W_{\imath',\emph{\textbf{j}}^{\:\prime}};\theta_0,\eta_0)\Big] \\
\lesssim& 
\Epn\Big[\Big\|\sumN\psi(W_{\imath,\emph{\textbf{j}}};\theta_0,\eta_0)\Big\|^2\Big]\le c_1^2.
\end{align*}
\noindent \textbf{Step 5.}
Note that all singular values of $J_0$ are bounded from above by $c_1$ under Assumption \ref{a:linear_orthogonal_score 2} (v) and all eigenvalues of $\Gamma$ are bounded from below by $c_0$ under Assumption \ref{a:regularity_nuisance_parameters 2} (iv). 
Therefore, we have $\|\sigma^{-1}\|\le c_1/\sqrt{c_0}$ and thus
$
\|\sigma^{-1}\|
=\Opn(1).
$
This completes a proof of (\ref{eq:step5}).
\end{proof}

\subsection{Proof of Theorem \ref{theorem:variance_estimator_linear 2}}\label{sec:theorem:variance_estimator_linear 2}
\begin{proof}
Step 2 of the proof of Theorem \ref{theorem:main_result_linear 2} proves $\|\hat J - J_0\| = O_p(\C^{-1/2} + r_n)$ and Assumption \ref{a:linear_orthogonal_score 2} (v) implies $\|J_0^{-1}\| \leq c_0^{-1}$.
Therefore, to prove the claim of the theorem, it suffices to show
\begin{align*}
& \ \Big\|\frac{1}{K^\ell}\sumk
\frac{|\underline{I_\kk}|}{\prodI}
\sum_{i=1}^\ell \sum_{\substack{\emph{\textbf{j}},\emph{\textbf{j}}^{\:\prime}\in I_\kk\\I_i(\emph{\textbf{j}})=I_i(\emph{\textbf{j}}^{\:\prime})}}
\sumN\sumjjj
\psi(W_{\imath,\emph{\textbf{j}}};\tilde \theta,\hat\eta_{\kk}) \psi(W_{\imath',\emph{\textbf{j}}^{\:\prime}};\tilde \theta,\hat\eta_{\kk})'
\\
&-
\sum_{i=1}^\ell\bar\mu_i\Ep\Big [
\sum_{\imath=1}^{N_\textbf{1}}\sum_{\imath'=1}^{N_{2_i}}
\psi(W_{\imath,\textbf{1}};\theta_0,\eta_0)\psi(W_{\imath',2_i};\theta_0,\eta_0)'\Big] \ \Big\|=\Op(\rho_n).
\end{align*}
Moreover, since $K$ and $d_{\theta}$ are constants and $\mu_i\to \bar \mu_i\le 1$, it suffices to show that for each $\kk\in [K]^\ell$ and $l,m\in[d_\theta]$, it holds that \small
\begin{align*}
\Big|
\frac{|\underline{I_\kk}|}{\prodI}\sum_{\substack{\emph{\textbf{j}},\emph{\textbf{j}}^{\:\prime}\in I_\kk\\I_i(\emph{\textbf{j}})=I_i(\emph{\textbf{j}}^{\:\prime})}}
\sumN
\sumjjj
\psi_l(W_{\imath,\emph{\textbf{j}}};\tilde \theta,\hat\eta_{\kk}) \psi_m(W_{\imath',\emph{\textbf{j}}^{\:\prime}};\tilde \theta,\hat\eta_{\kk})-\mu_i\Ep \Big [
\sum_{\imath=1}^{N_\textbf{1}}\sum_{\imath'=1}^{N_{2_i}}
\psi_l(W_{\imath,\textbf{1}};\theta_0,\eta_0)\psi_m(W_{\imath',2_i};\theta_0,\eta_0)\Big] \Big|
\\
=\Op(\rho_n).
\end{align*}\normalsize

Denote the left-hand side of the equation as $\mathcal I_{\kk,lm}$. 
First, note that $|\underline{I}|/|I_{k_i}|=\mu_i$.
We denote $i'$ for $I_{k_i}$ such that  $|I_{k_{i'}}|=|\underline{I_\kk}|$,
and apply the triangle inequality to get
\begin{align*}
\mathcal I_{\kk,lm}\le \mathcal I_{\kk,lm,1}+ \mathcal I_{\kk,lm,2},
\end{align*}
where
\begin{align*}
&\mathcal I_{\kk,lm,1}:=
\Big|\frac{1}{\prod_{i\neq i'}|I_{k_i}|^2 |I_{k_{i'}}|}\sum_{\substack{\emph{\textbf{j}},\emph{\textbf{j}}^{\:\prime}\in I_\kk\\I_i(\emph{\textbf{j}})=I_i(\emph{\textbf{j}}^{\:\prime})}}
\Big\{\sumN
\sumjjj\psi_l(W_{\imath,\emph{\textbf{j}}};\tilde \theta,\hat\eta_{\kk}) \psi_m(W_{\imath',\emph{\textbf{j}}^{\:\prime}};\tilde \theta,\hat\eta_{\kk})
\\
&\qquad\qquad\qquad\qquad\qquad\qquad\qquad\qquad
-
 \sumN
\sumjjj\psi_l(W_{\imath,\emph{\textbf{j}}};\theta_0,\eta_0) \psi_m(W_{\imath',\emph{\textbf{j}}^{\:\prime}};\theta_0,\eta_0) \Big\}
\Big|,\\
&\mathcal I_{\kk,lm,2}:=
\Big|\frac{1}{\prod_{i\neq i'}|I_{k_i}|^2 |I_{k_{i'}}|}\sum_{\substack{\emph{\textbf{j}},\emph{\textbf{j}}^{\:\prime}\in I_\kk\\I_i(\emph{\textbf{j}})=I_i(\emph{\textbf{j}}^{\:\prime})}} 
\sumN
\sumjjj\psi_l(W_{\imath,\emph{\textbf{j}}};\theta_0,\eta_0) \psi_m(W_{\imath',\emph{\textbf{j}}^{\:\prime}};\theta_0,\eta_0)
\\
&\qquad\qquad\qquad\qquad\qquad\qquad\qquad\qquad
-
\Ep [\sum_{\imath=1}^{N_\textbf{1}}\sum_{\imath'=1}^{N_{2_i}}\psi_l(W_{\imath,\textbf{1}};\theta_0,\eta_0)\psi_m(W_{\imath',2_i};\theta_0,\eta_0)]
\Big|.
\end{align*}
We first find a bound for $\mathcal I_{\kk,lm,2}$. Since $q>4$, it holds that
\begin{align*}
\Ep[\mathcal I_{\kk,lm,2}^2]
=&
\frac{1}{\prod_{i\neq i'} |I_{k_i}|^4|I_{k_{i'}}|^2}\Ep\Big[
\Big|\sum_{\substack{\emph{\textbf{j}},\emph{\textbf{j}}^{\:\prime}\in I_\kk\\I_i(\emph{\textbf{j}})=I_i(\emph{\textbf{j}}^{\:\prime})}} 
\sumN
\sumjjj
\psi_l(W_{\imath,\emph{\textbf{j}}};\theta_0,\eta_0) \psi_m(W_{\imath',\emph{\textbf{j}}^{\:\prime}};\theta_0,\eta_0)
\\
&\qquad\qquad\qquad\qquad\qquad
-
\Ep \Big[\sum_{\imath=1}^{N_\textbf{1}}\sum_{\imath'=1}^{N_{2_i}}\psi_l(W_{\imath,\textbf{1}};\theta_0,\eta_0)\psi_m(W_{\imath',2_i};\theta_0,\eta_0)\Big] 
\Big|^2
\Big]\\
\le &
\frac{1}{\prod_{i\neq i'} |I_{k_i}|^4|I_{k_{i'}}|^2}\Ep\Big[
\sum_{\substack{\emph{\textbf{j}},\emph{\textbf{j}}^{\:\prime},\emph{\textbf{j}}'',\emph{\textbf{j}}''' \in I_\kk\\I_i(\emph{\textbf{j}})=I_i(\emph{\textbf{j}}^{\:\prime}),I_i(\emph{\textbf{j}}'')=I_i(\emph{\textbf{j}}''')}}
\sum\limits_{\substack{I_s(\emph{\textbf{j}})=I_s(\emph{\textbf{j}}'')\\s\neq i}}
\sumN
\sumjjj
\sum_{\imath''\in [N_{\emph{\textbf{j}}''}]}
\sum_{\imath'''\in [N_{\emph{\textbf{j}}'''}]}
\\
&\qquad\qquad\qquad\qquad\qquad
 \psi_l(W_{\imath,\emph{\textbf{j}}};\theta_0,\eta_0) \psi_m(W_{\imath',\emph{\textbf{j}}'};\theta_0,\eta_0)
\psi_l(W_{\imath'',\emph{\textbf{j}''}};\theta_0,\eta_0) \psi_m(W_{\imath''',\emph{\textbf{j}'''}};\theta_0,\eta_0)
\Big]\\
&+
\frac{1}{\prod_{i\neq i'} |I_{k_i}|^4|I_{k_{i'}}|^2}\Ep\Big[
\sum_{\substack{\emph{\textbf{j}},\emph{\textbf{j}}^{\:\prime},\emph{\textbf{j}}'',\emph{\textbf{j}}''' \in I_\kk\\I_i(\emph{\textbf{j}})=I_i(\emph{\textbf{j}}^{\:\prime})=I_i(\emph{\textbf{j}}'')=
I_i(\emph{\textbf{j}}''')}}
\sumN
\sumjjj
\sum_{\imath''\in [N_{\emph{\textbf{j}}''}]}
\sum_{\imath'''\in [N_{\emph{\textbf{j}}'''}]}
\\
&\qquad\qquad\qquad\qquad\qquad
\psi_l(W_{\imath,\emph{\textbf{j}}};\theta_0,\eta_0) \psi_m(W_{\imath',\emph{\textbf{j}}'};\theta_0,\eta_0)\psi_l(W_{\imath'',\emph{\textbf{j}}''};\theta_0,\eta_0) \psi_m(W_{\imath''',\emph{\textbf{j}}'''};\theta_0,\eta_0)
\Big]\\
&+o(|\underline{I_\kk}|^{-1}) + 0\\
\lesssim&\frac{1}{|\underline{I_\kk}|}\Ep\Big[\Big\|\sumNN\psi(W_{\imath,\textbf{0}};\theta_0,\eta_0)\Big\|^4\Big]\lesssim c_1^4/\C=O(\C^{-1}).
\end{align*}\normalsize

Now, to bound $\mathcal I_{\kk,lm,1}$, we make use of the following identity coming from the proof of Theorem 3.2 in CCDDHNR (\citeyear{CCDDHNR18}): for any numbers $a$, $b$, $\delta a$, $\delta b$ such that
$|a|\vee |b|\le c$ and $|\delta a |\vee |\delta b| \le r$, it holds that 
$
|(a+\delta a)(b+\delta b)-ab|\le 2r( c+r).
$ Denote $\psi_{\emph{\textbf{j}},h}:=\psi_l(W_{\imath,\emph{\textbf{j}}};\theta_0,\eta_0)$ and $\hat\psi_{\emph{\textbf{j}},h}:=\psi_l(W_{\imath,\emph{\textbf{j}}};\tilde \theta,\hat\eta_{\kk})$ for $h\in\{l,m\}$ and apply the above identity with $a=\sumN\psi_{\emph{\textbf{j}},l}$, $b=\sumjjj\psi_{\emph{\textbf{j}}^{\:\prime},m}$, $a+\delta a =\sumN\hat\psi_{\emph{\textbf{j}},l}$, $b+ \delta b=\sumjjj\hat\psi_{\emph{\textbf{j}}^{\:\prime},m}$, $r=\Big|\sumN\hat\psi_{\emph{\textbf{j}},l}-\sumN\psi_{\emph{\textbf{j}},l}\Big|\vee \Big|\sumjjj\hat\psi_{\emph{\textbf{j}}^{\:\prime},m}-\sumjjj\psi_{\emph{\textbf{j}}^{\:\prime},m}\Big|$ and $c=\Big|\sumN\psi_{\emph{\textbf{j}},l}\Big|\vee\Big|\sumjjj\psi_{\emph{\textbf{j}}^{\:\prime},m}\Big|$.
Then
\begin{align*}
\mathcal I_{\kk,lm,1}=&
\Big|\frac{1}{\prod_{i\neq i'}|I_{k_i}|^2 |I_{k_{i'}}|}\sum_{\substack{\emph{\textbf{j}},\emph{\textbf{j}}^{\:\prime}\in I_\kk\\I_i(\emph{\textbf{j}})=I_i(\emph{\textbf{j}}^{\:\prime})}} \Big\{
\sumN\sumjjj\hat\psi_{\emph{\textbf{j}},l} \hat\psi_{\emph{\textbf{j}}^{\:\prime},m}
-
\sumN\sumjjj\psi_{\emph{\textbf{j}},l} \psi_{\emph{\textbf{j}}^{\:\prime},m}  \Big\}
\Big|\\
\le &
\frac{1}{\prod_{i\neq i'}|I_{k_i}|^2 |I_{k_{i'}}|}\sum_{\substack{\emph{\textbf{j}},\emph{\textbf{j}}^{\:\prime}\in I_\kk\\I_i(\emph{\textbf{j}})=I_i(\emph{\textbf{j}}^{\:\prime})}} \Big|
\sumN\sumjjj\hat\psi_{\emph{\textbf{j}},l} \hat\psi_{\emph{\textbf{j}}^{\:\prime},m}
-
\sumN\sumjjj\psi_{\emph{\textbf{j}},l} \psi_{\emph{\textbf{j}}^{\:\prime},m}\Big |\\
\le& 
\frac{2}{\prod_{i\neq i'}|I_{k_i}|^2 |I_{k_{i'}}|}\sum_{\substack{\emph{\textbf{j}},\emph{\textbf{j}}^{\:\prime}\in I_\kk\\I_i(\emph{\textbf{j}})=I_i(\emph{\textbf{j}}^{\:\prime})}}\Big (\Big|\sumN\hat\psi_{\emph{\textbf{j}},l}-\sumN\psi_{\emph{\textbf{j}},l}\Big|\vee \Big|\sumjjj\hat\psi_{\emph{\textbf{j}}^{\:\prime},m}-\sumjjj\psi_{\emph{\textbf{j}}^{\:\prime},m}\Big|\Big)
\\
& \times \Big(\Big|\sumN\psi_{\emph{\textbf{j}},l}\Big|\vee\Big|\sumjjj\psi_{\emph{\textbf{j}}^{\:\prime},m}\Big|
+\Big|\sumN\hat\psi_{\emph{\textbf{j}},l}-\sumN\psi_{\emph{\textbf{j}},l}\Big|\vee\Big |\sumjjj\hat\psi_{\emph{\textbf{j}}^{\:\prime},m}-\sumjjj\psi_{\emph{\textbf{j}'},m}\Big|\Big)\\
\le& 
\Big(\frac{2}{\prod_{i\neq i'}|I_{k_i}|^2 |I_{k_{i'}}|}\sum_{\substack{\emph{\textbf{j}},\emph{\textbf{j}}^{\:\prime}\in I_\kk\\I_i(\emph{\textbf{j}})=I_i(\emph{\textbf{j}}^{\:\prime})}}\Big |\sumN\hat\psi_{\emph{\textbf{j}},l}-\sumN\psi_{\emph{\textbf{j}},l}\Big|^2\vee\Big |\sumjjj\hat\psi_{\emph{\textbf{j}}^{\:\prime},m}-\sumjjj\psi_{\emph{\textbf{j}}^{\:\prime},m}\Big|^2\Big)^{1/2}\\
&\times
 \Big(\frac{2}{\prod_{i\neq i'}|I_{k_i}|^2 |I_{k_{i'}}|}\sum_{\substack{\emph{\textbf{j}},\emph{\textbf{j}}^{\:\prime}\in I_\kk\\I_i(\emph{\textbf{j}})=I_i(\emph{\textbf{j}}^{\:\prime})}}\Big\{\Big|\sumN\psi_{\emph{\textbf{j}},l}\Big|\vee\Big|\sumjjj\psi_{\emph{\textbf{j}}^{\:\prime},m}\Big|
\\&\qquad\qquad\qquad\qquad\qquad
+\Big|\sumN\hat\psi_{\emph{\textbf{j}},l}-\sumN\psi_{\emph{\textbf{j}},l}\Big|\vee\Big |\sumjjj\hat\psi_{\emph{\textbf{j}}^{\:\prime},m}-\sumjjj\psi_{\emph{\textbf{j}}^{\:\prime},m}\Big|\Big\}^2\Big)^{1/2}\\
\le&
\Big(\frac{2}{\prod_{i\neq i'}|I_{k_i}|^2 |I_{k_{i'}}|}\sum_{\substack{\emph{\textbf{j}},\emph{\textbf{j}}^{\:\prime}\in I_\kk\\I_i(\emph{\textbf{j}})=I_i(\emph{\textbf{j}}^{\:\prime})}}\Big|\sumN\hat\psi_{\emph{\textbf{j}},l}-\sumN\psi_{\emph{\textbf{j}},l}\Big|^2\vee\Big |\sumjjj\hat\psi_{\emph{\textbf{j}}^{\:\prime},m}-\sumjjj\psi_{\emph{\textbf{j}}^{\:\prime},m}\Big|^2\Big)^{1/2}\\
&\times\Big\{
 \Big(\frac{2}{\prod_{i\neq i'}|I_{k_i}|^2 |I_{k_{i'}}|}\sum_{\substack{\emph{\textbf{j}},\emph{\textbf{j}}^{\:\prime}\in I_\kk\\I_i(\emph{\textbf{j}})=I_i(\emph{\textbf{j}}^{\:\prime})}}\Big|\sumN\psi_{\emph{\textbf{j}},l}\Big|^2\vee\Big|\sumjjj\psi_{\emph{\textbf{j}}^{\:\prime},m}\Big|^2\Big)^{1/2} 
\\
&
+
  \Big(\frac{2}{\prod_{i\neq i'}|I_{k_i}|^2 |I_{k_{i'}}|}\sum_{\substack{\emph{\textbf{j}},\emph{\textbf{j}}^{\:\prime}\in I_\kk\\I_i(\emph{\textbf{j}})=I_i(\emph{\textbf{j}}^{\:\prime})}}\Big |\sumN\hat\psi_{\emph{\textbf{j}},l}-\sumN\psi_{\emph{\textbf{j}},l}\Big|^2\vee \Big|\sumjjj\hat\psi_{\emph{\textbf{j}}^{\:\prime},m}-\sumjjj\psi_{\emph{\textbf{j}}^{\:\prime},m}\Big|^2\Big)^{1/2}\Big\},
\end{align*}
where the second to the last inequality follows the Cauchy-Schwartz's inequality and Minkowski's inequality.
Notice that
\begin{align*}
&\sum_{\substack{\emph{\textbf{j}},\emph{\textbf{j}}^{\:\prime}\in I_\kk\\I_i(\emph{\textbf{j}})=I_i(\emph{\textbf{j}}^{\:\prime})}}\Big|\sumN\psi_{\emph{\textbf{j}},l}\Big|^2\vee\Big|\sumjjj\psi_{\emph{\textbf{j}}^{\:\prime},m}\Big|^2\le   \max_{1\leq i\leq \ell}\{|I_{k_i}|\}\sumjj \Big\|\sumN\psi(W_{\imath,\emph{\textbf{j}}};\theta_0,\eta_0)\Big\|^2,\\
&\sum_{\substack{\emph{\textbf{j}},\emph{\textbf{j}}^{\:\prime}\in I_\kk\\I_i(\emph{\textbf{j}})=I_i(\emph{\textbf{j}}^{\:\prime})}}\Big|\sumN\hat \psi_{\emph{\textbf{j}},l}-\sumN\psi_{\emph{\textbf{j}},l}\Big|^2\vee\Big|\sumjjj\hat \psi_{\emph{\textbf{j}}^{\:\prime},m}-\sumjjj\psi_{\emph{\textbf{j}}^{\:\prime},m}\Big|^2
\\
&\qquad\qquad \le \max_{1\leq i\leq \ell}\{|I_{k_i}|\}\sumjj\Big \|\sumN\psi(W_{\imath,\emph{\textbf{j}}};\tilde\theta,\hat \eta_{\kk})-\sumN\psi(W_{\imath,\emph{\textbf{j}}};\theta_0,\eta_0)\Big\|^2.
\end{align*}
Thus, the above bound for $\mathcal I_{\kk,lm,1}$ implies that
\begin{align*}
\mathcal I_{\kk,lm,1}^2
\lesssim&
R_n\times \Big(\frac{1}{\prodll}\sum_{\emph{\textbf{j}}\in I_\kk}
\Big\|\sumN \psi(W_{\imath,\emph{\textbf{j}}};\theta_0, \eta_0) 
\Big\|^2
+
R_n
\Big),
\end{align*}
where
\begin{align*}
R_n:=\frac{1}{\prodll}\sum_{\emph{\textbf{j}}\in I_\kk}
\Big\| \sumN\psi(W_{\imath,\emph{\textbf{j}}};\tilde\theta,\hat \eta_{\kk}) -\sumN \psi(W_{\imath,\emph{\textbf{j}}};\theta_0,\eta_0) 
\Big\|^2.
\end{align*}
Notice that 
\begin{align*}
\frac{1}{\prodll}\sum_{\emph{\textbf{j}}\in I_\kk}
\Big\|\sumN \psi(W_{\imath,\emph{\textbf{j}}};\theta_0,\eta_0) 
\Big\|^2=\Op(1),
\end{align*}
which is implied by Markov's inequality and the calculations
\begin{align*}
\Ep\Big[\frac{1}{\prodll}\sum_{\emph{\textbf{j}}\in I_\kk}
\Big\|\sumN \psi(W_{\imath,\emph{\textbf{j}}};\theta_0,\eta_0) 
\Big\|^2\Big]=&\Ep\Big[\Big \|\sum\limits_{\imath=1}^{N_{\textbf{0}}}\psi(W_{\imath,\textbf{0}};\theta_0,\eta_0) 
\Big\|^2\Big]\le c_1^2
\end{align*}
under Assumptions \ref{a:sampling 2} and \ref{a:regularity_nuisance_parameters 2} (ii). 
Finally, to bound $R_n$, using Assumption \ref{a:linear_orthogonal_score 2} (ii), 
\begin{align*}
R_n\lesssim&
 \frac{1}{\prodll}\sum_{\emph{\textbf{j}}\in I_\kk}
\Big\|\sumN \psi^a(W_{\imath,\emph{\textbf{j}}};\hat \eta_{\kk})(\tilde \theta -\theta_0)
\Big\|^2 
\\
&+
\frac{1}{\prodll}\sum_{\emph{\textbf{j}}\in I_\kk}
\Big\| \sumN\psi(W_{\imath,\emph{\textbf{j}}};\theta_0,\hat\eta_{\kk}) -\sumN\psi(W_{\imath,\emph{\textbf{j}}};\theta_0,\eta_0) 
\Big\|^2.
\end{align*}
The first term on RHS is bounded by
\begin{align*}
 \Big(\frac{1}{\prodll}\sum_{\emph{\textbf{j}}\in I_\kk}
\Big\|\sumN \psi^a(W_{\imath,\emph{\textbf{j}}};\hat \eta_{\kk})
\Big\|^2 \Big)\times\|\tilde \theta -\theta_0\|^2=\Op(1)\times \Op(\C^{-1})=\Op(\C^{-1})
\end{align*}
due to Assumption \ref{a:regularity_nuisance_parameters 2} (ii), Markov's inequality, and Theorem \ref{theorem:main_result_linear 2}.
Furthermore, given that $(W_{\imath,\emph{\textbf{j}}})_{\emph{\textbf{j}}\in I_{\kk}^c}$ satisfies $\hat \eta_{\kk}\in\mathcal T_n$,
\begin{align*}
&\Ep\Big[\Big \|\sumN\psi(W_{\imath,\emph{\textbf{j}}};\theta_0,\hat\eta_{\kk}) -\sumN\psi(W_{\imath,\emph{\textbf{j}}};\theta_0,\eta_0)\Big\|^2\Big|I_{\kk}^c\Big]
\\
\le&
\sup_{\eta\in \mathcal T_n}\Ep\Big[ \Big\|\sumN\psi(W_{\imath,\emph{\textbf{j}}};\theta_0,\eta) -\sumN\psi(W_{\imath,\emph{\textbf{j}}};\theta_0,\eta_0)\Big\|^2\Big|I_{\kk}^c\Big] \le (r_n')^2
\end{align*}
due to Assumptions \ref{a:sampling 2} and \ref{a:regularity_nuisance_parameters 2} (iii).
Also, the event $\hat \eta_{\kk}\in\mathcal T_n$ happens with probability $1-o(1)$, we have $R_n=\Op(\C^{-1}+(r'_n)^2)$. Thus we conclude that 
\begin{align*}
\mathcal I_{\kk,lm,1}=\Op(\C^{-1/2}+r'_n).
\end{align*}
This completes the proof.
\end{proof}

\section{Additional Lemma}
In this section, we establish a multiway generalization of Lemma \ref{lemma:hajek}.
For any $ r=1,...,\ell$,we let $\mathcal{I}_{r}(\textbf{C})=\Big\{\textbf{c}=\emph{\textbf{j}} \odot \textbf{e}: \textbf{e} \in \mathcal{E}_{r}, \mathbf{1} \leq \emph{\textbf{j}} \leq \textbf{C}\Big\}$ and $\boldsymbol{\varepsilon}_m=\{\textbf{e} \in\{0 ; 1\}^{\ell}: \sum_{i=1}^{\ell} e_{i}=m\}$ , with $\odot$ the Hadamard product on $\mathbb{R}^{\ell}$.

For each $n\in \mathbbm N$, let $(N^n_\emph{\textbf{j}},(W^n_{\imath,\emph{\textbf{j}}})_{1\le \iota \le \overline N})_{\emph{\textbf{j}}\geq \textbf{1}}$ be a set of random variables. For any $f:\supp(W^n)\to  \Real^d$ for a fixed $d\in \mathbbm N$, let us define the multiway empirical process
\begin{align*}
\GC f:=\sqrt{\C}\Big\{\frac{1}{\prod_C} \sum\limits_{i=1}^\ell\sum\limits_{j_i=1}^{C_i} \sum_{\iota \in N^n_\textbf{j}} f(W^n_{\imath,\emph{\textbf{j}}}) - \Ep[\sum_{\imath\in[N^n_\textbf{1}]}f(W^n_{\imath,\textbf{1}})]\Big\}.
\end{align*} 

\begin{lemma}[Independentization via H\'ajek Projections]\label{lemma:hajek_multiway}
For each $n\in \mathbbm N$, suppose that $(N^n_\emph{\textbf{j}},(W^n_{\imath,\emph{\textbf{j}}})_{1\le \iota \le \overline N})_{\emph{\textbf{j}}\geq \textbf{1}}$ satisfies Assumption \ref{a:sampling 2}. Let $\mathcal F_n$, $|\mathcal F_n|= d$, be a family of functions $f:\supp(W^n)\to \Real$ that satisfies $\mathbb{E}\Big[\Big(\sum\limits_{\imath\in [N^n_\textbf{1}]} f\Big(W^n_{\imath, \textbf{1}}\Big)\Big)^{2}\Big]<K<\infty$ for some $K$ independent of $n$. In addition, assume that $\underline{C}\rightarrow \infty$ and for every $\textbf{e} \in \boldsymbol{\varepsilon}_1$, $ \frac{\underline{C}}{\prod_C} \rightarrow \overline \mu_i \geq 0$, where $i$ is the nonzero coordinate of $\textbf{e}$. Then there exists a family of mutually independent standard uniform r.v.'s $(U_\textbf{c})_{\textbf{c}>0}$ such that the $H_nf$, the H\'ajek projection of $G_nf$ on the set of statistics of the form $\sum_{\textbf{c} \in \mathcal{I}_{r}(\textbf{C})} g_{\textbf{c}}(U_{\textbf{c}})$ (with $g_{\textbf{c}}(U_{\textbf{c}})$ square integrable, satisfies
\begin{equation}
H_{n} f=\sum_{\textbf{c} \in \mathcal{I}_{1}(\textbf{C})} \frac{\sqrt{\underline{C}}}{\prod_{i: \textbf{c}_{i} \neq 0} C_{i}}\left(\mathbb{E}\left[\sum_{\imath=1}^{N^n_{\textbf{c} \vee \textbf{1}}} f\left(W^n_{\imath, \textbf{c} \vee \textbf{1}}\right) \Big| U_{\textbf{c}}\right]-\mathbb{E}\left[\sum_{\imath \in [N^n_\textbf{1}]}f\left( {W}^n_{\imath,\textbf{1}}\right)\right]\right) .
\end{equation}
In addition, it holds uniformly over $\mathcal F_n$ that
\begin{align*}
V(\GC f)= V(H_n f)+O(\C^{-1})=\sum\limits_{\textbf{e}\in \boldsymbol{\varepsilon}_1}\bar\mu_i  Cov(\sum\limits_{\imath=1}^{N^n_\textbf{1}}f(W^n_{\imath,\textbf{1}}),\sum\limits_{\imath=1}^{N^n_{\textbf{2-\textbf{e}}}}f(W^n_{\imath,\textbf{2}-\textbf{e}})) +O(\C^{-1}).
\end{align*}
\begin{proof}
Throughout the proof, we drop the superscript $n$ for simplicity.
Under Assumption \ref{a:sampling 2}(i) and (ii), for each $n$, one can apply Lemma 7.35 of \cite{Kallenberg2006} and obtain  a measurable function $\tau_n$ such that
\begin{align}
(N_\emph{\textbf{j}},(W_{\imath,\emph{\textbf{j}}})_{1\le \iota \le \overline N})_{\emph{\textbf{j}}\geq \textbf{1}}=\big(\tau_n(U_{\emph{\textbf{j}}\odot\textbf{e}})_{\textbf{1}\prec \textbf{e}\preceq \textbf{1}}\big)_{\emph{\textbf{j}}\geq \textbf{1}}
\end{align}
where $(U_\textbf{c})_{\textbf{c}\geq \textbf{0}}$ denote a family of mutually independent uniform random variables on $[0,1]$. 

The rest of our proof closely follows that of Lemma D.2 in \cite{DDG18} with $r=\underline r=1$.
The H\'ajek projection $H_nf$ is characterized by 
\begin{align*}
\mathrm{E}\Big[\left(\mathbb{G}_{n} f-H_{n} f\right)\times \sum_{\textbf{c} \in \mathcal{I}_{1}(\textbf{C})} g_{\textbf{c}}\left(U_{\textbf{c}}\right)\Big]=0 \text{ for any } \left(g_{\textbf{c}}\right)_{\textbf{c} \in \mathcal{I}_{1}(\textbf{C})} \in\left(L^{\ell}([0 ; 1])\right)^{\left|\mathcal{I}_{1}(\textbf{C})\right|}.
\end{align*}
As a result,we have
\begin{equation*}
\mathrm{E}\left[\mathbb{G}_{n} f | U_{\textbf{c}}\right]=\mathrm{E}\left[H_{n} f | U_{\textbf{c}}\right] \text{ for any } \textbf{c}\in \mathcal{I}_1(\textbf{C}) .
\end{equation*}
Because the range $H_n$ is closed subspace of square integrable random variables,

\begin{equation*}
H_{n} f=\sum_{\textbf{c} \in \mathcal{I}_{1}(\textbf{C})} \mathbb{E}\left(H_{n} f | U_{\textbf{c}}\right) .
\end{equation*} 
Next
\begin{align*}
H_{n} f=\sum_{\textbf{c} \in \mathcal{I}_{1}(\textbf{C})} \mathbb{E}\left(\mathbb{G}_{n} f | U_{\textbf{c}}\right) .
\end{align*}
Note that for any \( \textbf{c} \in \mathcal{I}_{1}(\textbf{C})\), \(\textbf{c} \wedge \textbf{1}\) is the unique element $\boldsymbol{\varepsilon}_1$ such that $\textbf{c}=\emph{\textbf{j}} \odot \textbf{e}$ for some $\emph{\textbf{j}}$ (note that $\emph{\textbf{j}}$ is not unique).
Moreover, for any $\textbf{c} \in \mathcal{I}_{1}(\textbf{C})$ independence between the $U^{\prime} $ s ensures
that $\sumN  f\left(W_{\imath, \emph{\textbf{j}}}\right) \perp U_{\textbf{c}} \text { if } \emph{\textbf{j}} \odot \textbf{e} \neq \textbf{c}$. This implies
\begin{align*}
\mathbb{E}\left(\mathbb{G}_{n} f | U_{\textbf{c}}\right)
& =\frac{\sqrt{\underline{C}}}{\Pi_{C}} \sum_{\textbf{1} \leq \emph{\textbf{j}} \leq \textbf{C}} \mathbb{E}\left[\sumN f\left(W_{\imath, \emph{\textbf{j}}}\right)-\mathbb{E}\left[\sum_{\imath\in[N_\textbf{1}]}f\left(W_{\imath,\textbf{1}}\right)\right] \Big| U_{\textbf{c}}\right] \\
& =\frac{\sqrt{C}}{\Pi_{C}} \sum_{\textbf{1} \leq \emph{\textbf{j}} \leq \textbf{C}}           \mathds{1}  \{\emph{\textbf{j}} \odot \textbf{e}=\textbf{c}\} \mathbb{E}\left[\sumN f\left(W_{\imath, \emph{\textbf{j}}}\right)-\mathbb{E}\left[\sum_{\imath\in[N_\textbf{1}]}f\left( W_{\imath,\textbf{1}}\right)\right] \Big| U_{\textbf{c}}\right].
\end{align*}
The representation of $(N_\emph{\textbf{j}},(W_{\imath,\emph{\textbf{j}}})_{1\le \iota \le \overline N})_{\emph{\textbf{j}}\geq \textbf{1}}$ in terms of 
the $U$'s implies that 
\begin{equation*}
\mathbb{E}\left[\sum_{\imath=1}^{N_{\emph{\textbf{j}}}} f\left(W_{\imath, \emph{\textbf{j}}}\right)-\mathbb{E}\left[\sum_{\imath\in[N_\textbf{1}]}f\left(N_{\textbf{1}}, W_{\imath,\textbf{1}}\right)\right] \Big| U_{\textbf{c}}\right]=\mathbb{E}\left[\sum_{\imath=1}^{N_{\textbf{c} \vee \textbf{1}}} f\left(W_{\imath, \textbf{c} \vee \textbf{1}}\right)-\mathbb{E}\left[\sum_{\imath\in[N_\textbf{1}]}f\left( W_{\imath,\textbf{1}}\right)\right] \Big| U_{\textbf{c}}\right]
\end{equation*}
for any $ \emph{\textbf{j}}$ such that $\emph{\textbf{j}} \odot \textbf{e}=\textbf{c}$. Moreover,

\begin{align*}
\mathbb{E}\left(\mathbb{G}_{n} f | U_{\textbf{c}}\right)
& =\frac{\sqrt{\underline{C}}}{\Pi_{C}} \sum_{\textbf{1} \leq \emph{\textbf{j}} \leq \textbf{C}} \mathds{1}\{\emph{\textbf{j}} \odot \textbf{e}=\textbf{c}\} \mathbb{E}\left[\sum_{\imath=1}^{N_{ \textbf{c}\vee\textbf{1}}} f\left(W_{\imath, \textbf{c}\vee \textbf{1}}\right)-\mathbb{E}\left[\sum_{\imath\in[N_\textbf{1}]}f\left( W_{\imath,\textbf{1}}\right)\right] \Big| U_{\textbf{c}}\right] \\ 
& =\frac{\sqrt{\underline{C}} \prod_{i: \textbf{c}_{i}=0} C_{i}}{\Pi_{C}} \mathbb{E}\left[\sum_{\imath=1}^{N_{\textbf{c}\vee \textbf{1}}} f\left(W_{\imath, \textbf{c}\vee \textbf{1}}\right)-\mathbb{E}\left[\sum_{\imath\in[N_\textbf{1}]}f\left( W_{\imath,\textbf{1}}\right)\right] \Big| U_{\textbf{c}}\right]\\
& =\frac{\sqrt{\underline{C}}}{\prod_{i: \textbf{c}_{i} \neq 0} C_{i}}\left(\mathbb{E}\left[\sum_{\imath=1}^{N_{\textbf{c}\vee \textbf{1}}} f\left(W_{\imath, \textbf{c}\vee \textbf{1}}\right) \Big| U_{\textbf{c}}\right]-\mathbb{E}\left[\sum_{\imath\in[N_\textbf{1}]}f\left(W_{\imath,\textbf{1}}\right)\right]\right).
\end{align*}
It follows that
\begin{equation*}
H_{n} f=\sum_{\textbf{c} \in \mathcal{I}_{1}(\textbf{C})} \frac{\sqrt{\underline{C}}}{\prod_{i: \textbf{c}_{i} \neq 0} C_{i}}\left(\mathbb{E}\left[\sum_{\imath=1}^{N_{\textbf{c}\vee \textbf{1}}} f\left(W_{\imath, \textbf{c}\vee \textbf{1}}\right) \Big| U_{\textbf{c}}\right]-\mathbb{E}\left[\sum_{\imath\in[N_\textbf{1}]}f\left( W_{\imath,\textbf{1}}\right)\right]\right).
\end{equation*}
This shows the first claim of the lemma.

Since $\mathcal F_n$ is a finite family, we are left to prove that for each $f\in\mathcal F_n$,
\begin{align*}
V(\GC f)= V(H_n f)+O(\C^{-1})=\sum\limits_{i=1}^\ell\bar\mu_i  Cov(\sum\limits_{\imath=1}^{N_\textbf{1}}f(W_{\imath,\textbf{1}}),\sum\limits_{\imath=1}^{N_{2_i}}f(W_{\imath,2_i})) +O(\C^{-1}),
\end{align*}
where $2_i$ denotes the $\ell-$tuple vector with 2 in each entry but for 1 in the $i-$th entry.
Note that
\begin{equation}
\mathbb{V}\left(H_{n} f\right)=\sum_{e \in \boldsymbol{\varepsilon}_{1}} \frac{\underline{C}}{\prod_{i: \textbf{e}_{i}=1} C_{i}} \mathbb{V}\left(\mathbb{E}\left[\sum_{\imath \in [N_\textbf{1}]} f\left(W_{\imath, \textbf{1}}\right)\Big | U_{\textbf{e}}\right]\right).
\end{equation}
To conclude, it suffices to show that for each $\textbf{e}\in \boldsymbol{\varepsilon}_1$,
\begin{align*}
\mathbb{V}\left(\mathbb{E}\left[\sum_{\imath \in [N_\textbf{1}]} f\left(W_{\imath, \textbf{1}}\right)\Big | U_{\textbf{e}}\right]\right)=Cov\left(\sum_{\imath \in [N_\textbf{1}]} f\left(W_{\imath, \textbf{1}}\right), \sum_{\imath=1}^{N_{\textbf{2}-\textbf{e}}} f\left(W_{\imath, \textbf{2}-\textbf{e}}\right)\right).
\end{align*}
As $(N_\emph{\textbf{j}},(W_{\imath,\emph{\textbf{j}}})_{1\le \iota \le \overline N})_{\emph{\textbf{j}}\geq \textbf{1}} =
\left(\tau\left(\left(U_{\emph{\textbf{j}} \odot \textbf{e}}\right)_{\textbf{e} \in \cup_{r=1}^{\ell} \boldsymbol{\varepsilon}_{r}}\right)\right)_{\emph{\textbf{j}}\geq \textbf{1}} $ with i.i.d. $U$'s, we have 
$\mathbb{E}\left[\sum\limits_{\imath\in [N_\textbf{1}]} f\left(W_{\imath, \mathbf{1}}\right) \Big| U_{\textbf{e}}\right]=
\mathbb{E}\left[\sumN f\left(W_{\imath, \emph{\textbf{j}}}\right) \Big| U_{\textbf{e}}\right]$ for any
$\emph{\textbf{j}}$ such that $\emph{\textbf{j}} \odot \textbf{e}=\textbf{1} \odot \textbf{e}=\textbf{e}$. Becuase $\textbf{2}-\textbf{e} \odot \textbf{e}=\textbf{e}$, we have 
$\mathbb{V}\left(\mathbb{E}\left[\sum\limits_{\imath \in [N_\textbf{1}]} f\left(W_{\imath, \textbf{1}}\right)\Big | U_{\textbf{e}}\right]\right)
= Cov\left(\mathbb{E}\left[\sum\limits_{\imath \in [N_\textbf{1}]} f\left(W_{\imath, \textbf{1}}\right) \Big| U_{\textbf{e}}\right], \mathbb{E}\left[\sum\limits_{\imath=1}^{N_{\textbf{2}-\textbf{e}}} f\left(W_{\imath, \textbf{2}-\textbf{e}}\right)\Big | U_{\textbf{e}}\right]\right)$.
For any $\textbf{e} \in \boldsymbol{\varepsilon}_1$, we have $\textbf{2} - \textbf{e} \neq \textbf{1}$. The independence of the $U$'s ensures

\begin{equation*}
\left(U_{\textbf{1} \odot \textbf{e}^{\prime}}\right)_{\textbf{e}^{\prime} \in \cup_{r=1}^{\ell} \boldsymbol\varepsilon_{r} \backslash \textbf{e}} \bot\left(U_{(\textbf{2}-\textbf{e}) \odot \textbf{e}^{\prime}}\right)_{\textbf{e}^{\prime} \in \cup_{r=1}^{\ell} \boldsymbol\varepsilon_{r} \backslash \textbf{e}} | U_{\textbf{e}}
\end{equation*}

and thus $\sum\limits_{\imath=1}^{N_\textbf{1}} f\left(W_{\imath, \textbf{1}}\right) \perp \sum\limits_{\imath=1}^{N_{\textbf{2}-\textbf{e}}} f\left(W_{\imath, \textbf{2}-\textbf{e}}\right) | U_{\textbf{e}}$.

Hence, for $\textbf{e} \in \boldsymbol\varepsilon_1$
\begin{equation*}
\mathbb{E}\left[Cov\left(\sum\limits_{\imath\in[N_\textbf{1}]} f_{1}\left(W_{\imath, \textbf{1}}\right), \sum_{\imath=1}^{N_{\textbf{2}-\textbf{e}}} f_{2}\left(W_{\imath, \textbf{2}-\textbf{e}}\right)\Big | U_{\textbf{e}}\right)\right]=0.
\end{equation*}

By the law of total covariance, we obtain
\begin{equation*}
\mathbb{V}\left( \mathbb{E} \left[  \sum\limits_{\imath \in [N_\textbf{1}]} f (W_{\imath, \textbf{1}}) \Big| U_{\textbf{e}} \right] \right)
=Cov\left(\sum\limits_{\imath \in [N_\textbf{1}]}f\left(W_{\imath, \textbf{1}}\right), \sum_{\imath=1}^{N_{\textbf{2}-\textbf{e}}} f\left(W_{\imath, \textbf{2}-\textbf{e}}\right)\right) ~.
\end{equation*}
This establishes the second claim of the lemma.
\end{proof}
\end{lemma}

\newpage
\bibliography{draft_2020_03_04}

\newpage
\begin{table}[t]
	\centering
		\begin{tabular}{cccccrcccc}
		\hline\hline
			$N$ & $M$ & $\C$ & dim$(X)$ & $K$ ($K^2$) & Machine Learning & Bias & SD & RMSE & Cover \\
			\hline
			 25  &  25 &  25 & 100 & 2 \ (4) & Ridge       & 0.069 & 0.074 & 0.102 & 0.835\\
			     &     &     &     &         & Elastic Net & 0.010 & 0.079 & 0.080 & 0.963\\
			     &     &     &     &         & Lasso       & 0.005 & 0.080 & 0.080 & 0.965\\	
			\hline
			 50  &  50 &  50 & 100 & 2 \ (4) & Ridge       & 0.014 & 0.047 & 0.049 & 0.940\\
			     &     &     &     &         & Elastic Net &-0.002 & 0.048 & 0.048 & 0.956\\
			     &     &     &     &         & Lasso       &-0.001 & 0.049 & 0.049 & 0.955\\	
			\hline
			 25  &  25 &  25 & 200 & 2 \ (4) & Ridge       & 0.190 & 0.053 & 0.197 & 0.118\\
			     &     &     &     &         & Elastic Net & 0.016 & 0.077 & 0.079 & 0.969\\
			     &     &     &     &         & Lasso       & 0.006 & 0.080 & 0.080 & 0.968\\	
			\hline
			 50  &  50 &  50 & 200 & 2 \ (4) & Ridge       & 0.037 & 0.046 & 0.058 & 0.876\\
			     &     &     &     &         & Elastic Net &-0.000 & 0.048 & 0.048 & 0.960\\
			     &     &     &     &         & Lasso       &-0.002 & 0.048 & 0.048 & 0.962\\	
			\hline
			 25  &  25 &  25 & 100 & 3 \ (9) & Ridge       & 0.042 & 0.074 & 0.085 & 0.962\\
			     &     &     &     &         & Elastic Net & 0.004 & 0.074 & 0.074 & 0.993\\
			     &     &     &     &         & Lasso       & 0.002 & 0.075 & 0.075 & 0.992\\	
			\hline
			 50  &  50 &  50 & 100 & 3 \ (9) & Ridge       & 0.007 & 0.048 & 0.049 & 0.962\\
			     &     &     &     &         & Elastic Net &-0.001 & 0.047 & 0.047 & 0.972\\
			     &     &     &     &         & Lasso       &-0.001 & 0.048 & 0.048 & 0.963\\	
			\hline
			 25  &  25 &  25 & 200 & 3 \ (9) & Ridge       & 0.081 & 0.067 & 0.105 & 0.896\\
			     &     &     &     &         & Elastic Net & 0.005 & 0.073 & 0.073 & 0.994\\
			     &     &     &     &         & Lasso       & 0.003 & 0.076 & 0.077 & 0.992\\	
			\hline
			 50  &  50 &  50 & 200 & 3 \ (9) & Ridge       & 0.018 & 0.047 & 0.050 & 0.944\\
			     &     &     &     &         & Elastic Net &-0.002 & 0.048 & 0.048 & 0.968\\
			     &     &     &     &         & Lasso       &-0.003 & 0.049 & 0.049 & 0.968\\	
		\hline\hline
		\end{tabular}
	\caption{Simulation results based on 5,000 Monte Carlo iterations. Results are displayed for each of the three machine learning methods, including the ridge, elastic net, and lasso. Reported statistics are the bias (Bias), standard deviation (SD), root mean square error (RMSE), and coverage frequency for the nominal probability of 95\% (Cover).}
	\label{tab:simulation_results}
\end{table}
\begin{table}
	\centering
	\scalebox{1}{
		\begin{tabular}{lcccccc}
			\hline\hline
			&& 0-Way & 1-Way & 1-Way & 2-Way\\
			Instrument $(Z_{ij})$ && --- & {\small Product} & {\small Market} & $\stackrel{\text{\scriptsize Product}}{\text{\scriptsize $\times$Market}}$\\
			\hline
			Horsepower/weight &&-5.763 &-5.719 &-5.815 &-5.659\\
			of other products &&(0.460)&(0.640)&(1.024)&(1.211)\\
			\hline
			Miles/dollar      &&-6.121 &-6.056 &-6.191 &-6.121\\
			of other products &&(0.607)&(0.865)&(1.491)&(3.963)\\
			\hline
			Size              &&-5.684 &-5.641 &-5.727 &-5.593\\
			of other products &&(0.413)&(0.565)&(0.892)&(1.015)\\
			\hline\hline
		\end{tabular}
	}
	\caption{Estimates and standard errors of the coefficient $\theta_0$ of log price in the demand model. The first column indicates the instrumental variable. The second column shows the results of the DML by lasso not accounting for clustering with the number $K=4$ of folds for cross fitting. The third and fourth columns show the results of the 1-way cluster-robust DML by lasso clustered at product and market, respectively, with the number $K=4$ of folds for cross fitting. The fifth column shows the results of the 2-way cluster-robust DML by lasso with the number $K^2=4$ of folds for two-way cross fitting. All the results are based on the average of ten rerandomized DML.}
	\label{tab:empirical_results}
\end{table}

\end{document}